\documentclass[11pt,preprint]{elsarticle}

\usepackage{hyperref}
\usepackage{setspace}
\usepackage[table]{xcolor}
\journal{}

\usepackage{bm}

\usepackage{dsfont}
\usepackage{array}              
 \usepackage{bm} 
 \usepackage{boldline}  
 \usepackage{multirow}           
 \usepackage{float}               
 \usepackage{enumitem}          
 \usepackage{ulem} 
 \usepackage{color}
 \usepackage{tikz}
\usepackage{algorithm}
\usepackage{amsthm}
\usepackage{comment}
\usepackage[noend]{algorithmic} 

\usepackage{stfloats}
\usepackage{subcaption}
\usepackage{multicol}
\usepackage{amsfonts}
\usepackage{mathtools}
\usepackage{arydshln}

\newcommand{\lV}{\vspace{-1mm}}
\newcommand{\llV}{\vspace{-2mm}}
\newcommand{\lllV}{\vspace{-5mm}}
\newcommand{\lH}{\hspace{-1mm}}

\newcommand{\mH}{\hspace{1mm}}
\newcommand{\mmH}{\hspace{2mm}}
\newcommand{\mmmH}{\hspace{5mm}}
\newcommand{\mV}{\vspace{1mm}}
\newcommand{\mmV}{\vspace{2mm}}

\newcommand{\bla}{\color{black}}

\newcommand{\mi}{-}
\newcommand{\ma}{+}

\newcommand{\av}{\bm{a}}
\newcommand{\bv}{\bm{b}}
\newcommand{\cv}{\bm{c}}

\newcommand{\rv}{\bm{r}}

\newcommand{\rvI}[1]{\bm{r}^{\mi1}_0(#1)}
\newcommand{\avh}{\bar\av}
\newcommand{\bvh}{\bar\bv}
\newcommand{\avt}{\underbar{$\av$}}

\renewcommand{\t}{\tau}
\newcommand{\tx}{\t_x}
\newcommand{\txS}{\hat\t_x}
\newcommand{\ks}{d_{KS}}
\newcommand{\ksP}{d_{P-KS}}
\newcommand{\ksN}{d_{NP-KS}}

\newcommand{\sA}{{V_{\av}}}              
\newcommand{\sB}{{V_{\bv}}}                
\newcommand{\sC}{{V_{\cv}}}                
\newcommand{\nr}{\bullet}          

\newcommand{\sign}{\operatorname{sign}}
\newcommand{\abs}[1]{\left\lvert#1\right\rvert}
\newcommand{\inter}{\bigcap} 
\newcommand{\AB}{\sA\inter\sB}
\newcommand{\ap}{\av|_{(\sA\inter \sB)}}
\newcommand{\bp}{\bv|_{(\sA\inter \sB)}}
\newcommand{\pte}{|_{V_{\av,\bv,\cv}}} 
\newcommand{\eproof}{$\;_\square$}

\newtheorem{definition}{Definition}
\newtheorem{theorem}{Theorem}
\newtheorem{corollary}{Corollary}
\newtheorem{axiom}{Axiom}

\newtheorem{innercustomcorollary}{Corollary}
\newenvironment{customcorollary}[1]
{\renewcommand\theinnercustomcorollary{#1}\innercustomcorollary}
{\endinnercustomcorollary}

\newtheorem{innercustomtheorem}{Theorem}
\newenvironment{customtheorem}[1]
{\renewcommand\theinnercustomtheorem{#1}\innercustomtheorem}
{\endinnercustomtheorem}

\begin{document}
\begin{titlepage}
    \begin{center}
        \vspace*{1cm}
        
        {\Large\textbf{A New Correlation Coefficient for Comparing and Aggregating Non-strict and Incomplete Rankings}}
        
        \vspace{0.5cm}
        
        \vspace{1.5cm}
        
        \textbf{Yeawon Yoo \footnote{        Corresponding author\\
        E-Mail: yyoo12@asu.edu\\
        Phone: +1-(480)-965-5248
}, Adolfo R. Escobedo, and J. Kyle Skolfield}

        \vspace{0.5cm}
        P.O. Box 878809, Tempe, AZ 85287-8809 \\ School of Computing, Informatics, and Decision Systems Engineering\\ Arizona State University\\
        \vspace{0.5cm}
        \{yyoo12, adres, joshua.skolfield\}@asu.edu

        \vfill
       
        \vspace{0.8cm}
    \end{center}

\end{titlepage}

\begin{frontmatter}

\title{A New Correlation Coefficient for Comparing and Aggregating Non-strict and Incomplete Rankings\lllV\lllV}

\begin{abstract}
We introduce a correlation coefficient that is designed to deal with a variety of ranking formats including those containing non-strict (i.e., with-ties) and incomplete (i.e., unknown) preferences. The correlation coefficient is designed to enforce a neutral treatment of incompleteness whereby no assumptions are made about individual preferences involving unranked objects. The new measure, which can be regarded as a generalization of the seminal Kendall tau correlation coefficient, is proven to satisfy a set of metric-like axioms and to be equivalent to a recently developed ranking distance function associated with Kemeny aggregation. In an effort to further unify and enhance both robust ranking methodologies, this work proves the equivalence of an additional distance and correlation-coefficient pairing in the space of non-strict incomplete rankings. These connections induce new exact optimization methodologies: a specialized branch and bound algorithm and an exact integer programming formulation. Moreover, the bridging of these complementary theories reinforces the singular suitability of the featured correlation coefficient to solve the general consensus ranking problem. The latter premise is bolstered by an accompanying set of experiments on random instances, which are generated via a herein developed sampling technique connected with the classic Mallows distribution of ranking data. Associated experiments with the branch and bound algorithm demonstrate that, as data becomes noisier, the featured correlation coefficient yields relatively fewer alternative optimal solutions and that the aggregate rankings tend to be closer to an underlying ground truth shared by a majority. 
\end{abstract}


\end{frontmatter}

\section{Introduction\llV}
The consensus ranking problem (i.e., ranking aggregation) is at the center of many group decision-making processes. It entails finding an ordinal vector or \textit{ranking} of a set of competing objects that minimizes disagreement with a profile of preferences (represented as ranking vectors). Common examples include corporate project selection, research funding processes, and academic program rankings \citep{hoc06met}. Moreover, the mathematical measures and aggregation algorithms devised to solve the consensus ranking problem often find ready application in many other fields. In Information Retrieval, these fundamental tools have been used to compare, aggregate, and evaluate the accuracy of metasearch engine lists \cite{hassanzadeh2014axiomatic}. \bla In recommendation systems, they have been used to evaluate the similarity between the characteristics of users or items. Based on the measured similarity, for example, the recommendation system suggests items that the most similar users have liked, or the most similar item that a user has liked  \cite{grzegorzewski2015recommender,lkacka2014measuring}. \bla Hence, although this work considers the group decision-making context of ranking aggregation for ease of interpretability, many of its results could be readily adapted to various other contexts such as Artificial Intelligence \citep{bet14the} and Biostatistics \cite{bruno2013microclan,burkovski2014rank,cohen2011using}.

Although the mathematical roots of consensus ranking trace back to the development of voting systems of de Borda \citep{de1781memoire} and Condorcet \citep{con85mar}, significant work remains to deal with real-world situations that upend many of the problem's rigid and long-running assumptions. These issues have garnered renewed interest from the Operations Research community owing largely to the general intractability engendered by the more robust ranking aggregation systems \citep{sen14ope}. Even so, it may be beneficial to consider transdisciplinary efforts in solving close variants of this problem. This work incorporates concepts from the statistical literature where the analogous \textit{median ranking problem} has been used for classification, prediction, and several other applications \citep{amb16are,hei13clu,mesiarova2012multi}. The fundamental goal of the present work for intertwining these viewpoints is to reinforce and advance theoretical and computational aspects of consensus ranking when dealing with indispensable forms of ranking data. Additionally, this unison is intended to yield insights and perspectives that are generalizable to various other contexts. 

This work deals with the consensus ranking problem in which the set of input rankings may contain ties and may be incomplete, which is the rule rather than the exception in group decision-making \citep{emo02new, kim1999interactive}. Hence, utilizing frameworks that possess this flexibility of preference expression is imperative;  otherwise, judges are implicitly forced to make arbitrary and/or careless choices. In essence, since there is typically a finite budget or set of benefits that is to be allocated commensurate with the competitors' positions in the consensus ranking, the chosen frameworks must employ robust measures that align with the given context. To be precise, this work adopts a \textit{neutral treatment} of incomplete rankings, whereby a judge's preferences over \bla his/her \bla unranked objects are unknown---because in the most general case it is assumed \bla he/she \bla does not evaluate them. Therefore, no inferences are made about individual preferences of unranked objects relative to other unranked or ranked objects. Such a treatment is particularly prudent for situations where the subjective evaluation of a large object set can be realistically accomplished only via the allocation of smaller subsets---which may differ both in content and size---to various judges. \bla Furthermore, this work emphasizes  desired social choice properties in group decision-making including the ability of the aggregate outcome to assign equitable electoral power to each ranking input and to reflect the preferences of a majority \citep{bra16han}.\bla

This work makes the following novel contributions. First, it demonstrates that the $\tx$ ranking correlation coefficient devised in \cite{emo02new} is inadequate for enforcing a neutral treatment of incomplete rankings. Second, it derives the $\txS$ ranking correlation coefficient and formally establishes its axiomatic foundation for dealing with a wide variety of ranking inputs. 
Third, it proves that the featured correlation coefficient---itself a generalization of the $\tau$ \citep{ken38new} and $\tx$ \citep{emo02new} correlation coefficients---is equivalent to an axiomatic ranking distance recently developed in \cite{mor16axi}, and it establishes a connection between another distance-correlation pairing. Altogether, the first three contributions refine and unify distance and correlation-based ranking aggregation. Fourth, it leverages these connections to induce new exact optimization methodologies for solving the non-strict incomplete ranking aggregation problem. Fifth, it extends the repeated insertion model \cite{doi04rep} to sample non-strict incomplete rankings from statistical distributions linked with the classic Mallows-$\phi$ distribution \cite{mal57non}; these instances are leveraged to evaluate the abilities of $\tx$ and $\txS$ to satisfy two key social choice properties.

The paper is organized as follows. \S\ref{Sec:Prelim} introduces the adopted notational conventions. \S\ref{Sec:Lit} reviews the pertinent literature on axiomatic distances and correlation coefficients for measuring differences and similarities between rankings. \S\ref{Sec:IRviaCC} derives theoretical results that strengthen the correlation-based framework for handling non-strict incomplete rankings and, in particular, its advantages to engender new exact optimization methodologies for the related ranking aggregation problem. \S\ref{Sec:Algorithms} develops an efficient statistical sampling framework from which nontrivial ranking aggregation instances are constructed; these are then used to compare key social choice-related properties associated with $\tx$ and $\txS$. Lastly, \S\ref{Sec:Discussion} concludes the work and discusses future avenues of research.\llV

\section{Notation and preliminary conventions\llV}\label{Sec:Prelim}
Denoting $V=\{v_1,\dots, v_n\}$ as a set of objects, a judge's \textit{ranking} or ordinal evaluation of $V$ is characterized by a vector $\av$ of dimension of $n$, whose $i$-th element denotes the ordinal position assigned to object $v_i$. If $a_i < a_j$, $\av$ is said to \textit{prefer} $v_i$ to $v_j$ (or to \textit{disprefer} $v_j$ to $v_i$), and when $a_i=a_j$, $\av$ is said to \textit{tie} $v_i$ and $v_j$, where $1\le i,j\le n$ and $i\neq j$. Additionally, when $a_i$ is assigned the null value ``$\bullet$'', $v_i$ is said to be unranked within $\av$; the objects explicitly ranked in $\av$ are denoted by the subset $V_{\av}\subseteq V$ (i.e., $a_i\ne\bullet$ for $v_i\in V_{\av}$). For example, in the $5$-object ranking $\av=(1,2,2,\bullet,4), v_1$ is preferred over $v_2,v_3,$ and $v_5$; $v_2$ and $v_3$ are tied for the second position but both are preferred over $v_5$; $v_4$ is left unranked; and $V_{\av}=V\backslash\{v_4\}$. It is important to emphasize that, according to the neutral treatment of incomplete rankings adopted in this work, although any object $v_i\in V$ that is unranked within $\av$ receives the same assignment $a_i=\bullet$, it is not considered tied with other unranked objects or better/worse than the ranked objects

Various ranking aggregation systems may not be defined or equipped to properly handle the full variety of ranking data formats alluded to in the preceding paragraphs. For this reason, the following definitions highlight three primary \textit{ranking spaces} by which they can be categorized.

\begin{definition}\label{def:univSpace}
Let $\Omega=\{\bullet,1,\dots,n\}^n$ denote the broadest ranking space consisting of all \textit{(\romannumeral 1) strict, (\romannumeral 2) non-strict, (\romannumeral 3) complete, and (\romannumeral 4) incomplete} rankings---corresponding to rankings (\romannumeral 1) without ties, (\romannumeral 2) with and without ties, (\romannumeral 3) full, and (\romannumeral 4) partial and full, respectively. Since non-strict and incomplete rankings also encompass strict and complete rankings, respectively, $\Omega$ is denoted henceforth as \textit{the space of non-strict incomplete rankings}.
\end{definition}

\begin{definition}\label{def:CSpace}
Let $\Omega_C=\{1,\dots,n\}^n$ denote \textit{the space of complete rankings} over $n$ objects, which consists of all non-strict (and strict) rankings where every object is explicitly ranked (i.e., partial evaluations are disallowed).
\end{definition}

\begin{definition}\label{def:SSpace}
Let $\Omega_S=\{\bullet,1,\dots,n\}^n$ denote \textit{the space of strict rankings} over $n$ objects, which consists of all incomplete (and complete) rankings where no objects are tied.
\end{definition}

From the above definitions, it is evident that $\Omega_C\subset\Omega$, $\Omega_S\subset\Omega$, and $\Omega_C$ and $\Omega_S$ are incomparable. To describe the ranking aggregation problem addressed in this work, let $\dot \t(\cdot):\Omega^2 \rightarrow [-1,1]^1$ denote an arbitrary ranking correlation function---here, the overhead dot is used to disassociate this general function from the standard Kendall-tau coefficient \citep{ken38new}, with which the undotted symbol is typically associated. \bla The correlation-based \textit{non-strict incomplete ranking aggregation problem (NIRA)} is stated formally as:\llV
\begin{equation}
   \arg\max_{\rv\in\Omega_C} \sum_{k=1}^{K}\dot\t(\rv,\av^k)\label{eqn:c-consensus},\lV
\end{equation}
where $\av^k\in\Omega$ for $k=1,\dots,K$ (i.e., $k$ is the index of each judge or ranking). Alternatively, letting $\dot d(\cdot):\Omega^2 \rightarrow \mathbb{R}^1_{+\cup \{0\}}$ denote an arbitrary ranking distance function\bla---the overhead dot disassociates this symbol from any specific distance function---\bla the distance-based NIRA is stated formally as:\llV
\begin{equation}
   \arg\min_{\rv\in\Omega_C} \sum_{k=1}^{K}\dot d(\rv,\av^k)\label{eqn:d-consensus}.\llV
\end{equation}
Expression \eqref{eqn:c-consensus} can be intuitively interpreted as the problem of finding a ranking $\rv$ that maximizes agreement---quantified according to $\dot\t$---collectively with $K$ non-strict incomplete rankings; Expression \eqref{eqn:d-consensus} can be intuitively interpreted as the problem of finding a ranking $\rv$ that minimizes disagreement---quantified according to $\dot d$---collectively with the same inputs. \bla For certain \bla distance and correlation-coefficient pairings (see \S\ref{SS:pairings}), the two respective optimization problems are equivalent. It is imperative to point out that, although the input rankings are allowed to be incomplete to allow flexibility of preference expression, the consensus ranking is required to lie in the space of complete rankings---that is, $\rv\in\Omega_C$ is a constraint of both problems.\llV 

\section{Literature review \llV}\label{Sec:Lit}
The principal focus of this work is on deterministic metric-based methods for comparing and aggregating rankings, which are regarded as the most robust methodologies within Operation Research and Social Choice \citep{bra16han}. The reader is directed to \cite{coo06dis} for a review of score or utility based methods, which are more computationally efficient but cannot fulfill certain fundamental properties associated with voting fairness (e.g., the Condorcet criterion \citep{con85mar} and its extensions \citep{you88con,you78con}). Additionally, there is a rich body of literature on nondeterministic or model-based ranking aggregation methods (e.g., see \citep{fli86dis,mal57non,mar14ana}). While these often rely on axiomatic distances, they are incomparable with the featured context in various notable respects including their assumptions, aggregation processes, and outputs. \llV

\subsection{Axiomatic distances}\label{SS:Distances}
Ranking aggregation has been primarily associated with axiomatic distances, several of which have been proposed to address different variants of the original problem \citep{coo06dis}. In these works, a distance function is typically advocated as the most suitable for aggregating inputs drawn from a specific ranking space through a set of mathematical axioms it uniquely satisfies. \bla Additional criteria for judging the suitability of different distances is through the verification of the social choice properties they satisfy; the latter can be described through said axioms (characterized by comparisons of two rankings) and through characteristics exhibited by the consensus rankings (characterized by comparisons with all of the input rankings).\bla

Among various notable axiomatic distances that have been introduced, only those proposed in \citep{coo07cre,dwo01ran,mor16axi} were purposely designed to deal with the broadest ranking space $\Omega$ under a neutral treatment of incomplete rankings. Hence, this subsection restricts most of its attention to these distances and on the precursor distance upon which they are founded. This first axiomatic distance was introduced by Kemeny and Snell in \cite{kem62pre} for rankings in $\Omega_C$; the distance function, written here succinctly as $\ks$, quantifies the disagreement between a pair of complete rankings as follows: \lV
\begin{align}
\ks(\av,\bv)&=\frac{1}{\gamma}\sum_{i=1}^n\sum_{j=1}^n\abs{\sign(a_i-a_j)-\sign(b_i-b_j)}, \lllV\label{eqn:ks}
\end{align}
where $\av,\bv\in\Omega_C$ and $\gamma$ is a constant associated with a chosen minimum positive distance unit. In \cite{kem62pre}, $\gamma=2$, corresponding to a minimum distance unit of 1 (since each object pair is counted twice in the above expression), but henceforth $\gamma$ is fixed to 4 corresponding to a minimum positive distance unit of 1/2, which does not affect the solution to Problem \eqref{eqn:d-consensus} but has a convenient interpretation for handling ties \citep{mor16axi}. Put simply, $\ks(\av,\bv)$ measures the number of pairwise rank reversals required to turn $\av$ into $\bv$. The distance is synonymous with robust ranking aggregation in space $\Omega_C$ \citep{ail08agg} owing to 
the combination of social choice properties it uniquely satisfies (see  \cite{you88con,you78con}).

Distance $\ks$ was extended to handle incomplete rankings in \citep{coo07cre,dwo01ran}. The underlying axioms for this distance, referred to as the \textit{Projected Kemeny Snell distance} and written here succinctly as $\ksP$, are provided in \cite{mor16axi}. The corresponding distance function between $\av,\bv\in\Omega$ is defined as:\llV
\begin{align}
\ksP(\av,\bv)=\ks(\ap,\bp),\label{eqn:ksP} \lllV
\end{align}
where $\ap$, $\bp$ denote the projections of each ranking onto the subset of objects \textit{evaluated in both rankings}. In other words, $\ksP$ enforces the intuitive interpretation that ranking disagreements should be based only on the objects ranked in common by $\av$ and $\bv$. 
That said, evidence from \citep{mor16axi} suggests that utilizing $\ksP$ may be undesirable for the group decision-making context due to an associated systematic bias. Therein, it was shown that despite the aligned preferences of a large majority, a few judges with opposing preferences can dominate the resulting consensus ranking by simply evaluating more objects. 

The normalized projected Kemeny Snell distance, written here succinctly as $\ksN$, was developed in \citep{mor16axi} to overcome the aforementioned drawback of $\ksP$. The $\ksN$ distance is equivalent to $\ks$ when the inputs are restricted to space $\Omega_C$, but it uniquely satisfies an intuitive set of axioms desired of any distance defined in space $\Omega$. The corresponding distance function between $\av,\bv\in\Omega$ is defined as:
\begin{equation}\label{eqn:npks}\small
\ksN(\av,\bv)=
\begin{cases}
\frac{\ks(\ap,\bp)}{\bar n(\bar n-1)/2}
& \text {if } \bar n \geq 2,\lV\\
0 & \text{otherwise},
\end{cases}\lV
\end{equation}
where $\bar n:=\abs{\AB}$. From its axiomatic foundation, $\ksN$ gives equitable voting power to each input ranking or judge in the aggregation process.  

\bla Before proceeding, it is important to briefly review other distances defined for incomplete rankings that do not conform with a neutral treatment of incompleteness herein adopted. Expressly, these measures make explicit or implicit assumptions about individual preferences over unevaluated objects. For instance, top-$k$ ranking distances explicitly assume that an individual's $(n-k)$ unranked objects are all tied for ordinal position $k\ma1$, making them all strictly dispreferred to the $k$ ranked objects (e.g., see \cite{kle08uns,mam07eff}). As an additional example, a set of axioms and logical conditions were defined in \cite{roy1993criterion} for a distance that assigns numeric values to four preferential relationships: \textit{preferred, dispreferred, indifferent,} and \textit{incomparable}. The distance was extended to solve the aggregation problem in \cite{jab04dis,khelifa2001distance}. Given two rankings, the distance value is $5/3$ when one ranking does not evaluate objects $v_i, v_j$ and another prefers $v_i$ over $v_j$ (or vice versa); if one ranking prefers $v_i$ over $v_j$ and the other one prefers $v_j$ over $v_i$, the distance value is 2. This is to say that the former has 5/6 the weight of the latter implicitly (the corresponding $\ksP$ or $\ksN$ distance values are not comparable). Readers are directed to \cite{bog73pre,bog75pre,coo86inf,kim16eff} for additional examples.\bla

\subsection{Correlation coefficients}
Correlation coefficients are an alternative methodology for comparing rankings with an extensive history and wide array of applications in statistical literature (e.g., \cite{koc87com,leh66som,liu12hig,opara2016computation}). They have been more recently applied to ranking aggregation \cite{amo16acc,emo02new}, where the agreement between judges $\av$ and $\bv$ is measured on the interval $[-1,1]$; the minimum/maximum values indicate complete disagreement/agreement. The most prominent is the Kendall $\t$ (tau) correlation coefficient \cite{ken38new}, which is applicable for rankings in space $\Omega_C\cap\Omega_S$. In \cite{ken48ran}, it was expanded to rankings in $\Omega_C$ as the $\t_b$ correlation coefficient. However, \citeauthor{emo02new} \cite{emo02new} gave compelling evidence that $\tau_b$ exhibits serious flaws including returning the undefined correlation of $0/0$ when comparing the all-ties ranking to itself or to any other non-strict ranking. To replace it, the authors introduced the $\tx$ (tau-extended) correlation coefficient, which relies on the following \textit{ranking-matrix} representation of $\av\in\Omega_C$, denoted as $[a_{ij}]$, with individual entries $1\leq i,j\leq n$ given by:\llV 
\begin{equation}\label{eqn:taux_sm}
a_{ij}=\left\{\begin{array}{rl} 1 &\mmH \text{if } \mH a_i\leq a_j,\\
-1 &\mmH \text{if } \mH a_i > a_j,\\
0 &\mmH \text{if } \mH i=j.\\
\end{array}\right.\llV\lV
\end{equation}
Here, a tie connotes a positive statement of agreement; conversely, the $\t_b$ ranking-matrix (not shown) treats a tie as a declaration of indifference by giving it a value of zero \citep{emo02new}. Applying Equation \eqref{eqn:taux_sm}, the $\tx$ correlation between $\av,\bv\in\Omega_C$, with respective ranking-matrices $[a_{ij}]$ and $[b_{ij}]$, is given by the function:
\begin{equation}\label{eqn:taux}
   \tx(\av,\bv) = \frac{\sum_{i=1}^{n}\sum_{j=1}^{n}a_{ij}b_{ij}}{n(n\mi1)}.
\end{equation}
In \cite{emo02new}, it was also proved that $\tx$ is connected to $\ks$ via the equation:\llV
\begin{equation}\label{eqn:taux_to_ks}
\tx(\av,\bv)=1-\frac{\gamma\;\ks(\av,\bv)}{n(n\mi1)},
\end{equation}
where $\gamma>0$ is the minimum $\ks$ distance unit (see Equation \eqref{eqn:ks}). This connection renders $\tx$ with a similar axiomatic foundation as $\ks$. At the same time, since the corresponding NIRA problems are also equivalent, it suggests that the inadequacies of the latter to handle incomplete rankings \cite{mor16axi} carry over to the former. The ensuing section will explore this premise. 

\bla We remark that alternative correlation coefficients have been defined for the space of incomplete rankings using concepts from fuzzy set theory, which deals with the representation of incomplete or vague information. In this context, missing ranking values are expressed as an interval \cite{slowinski2012fuzzy}---which serves to estimate the missing or incomparable information. This treatment is useful in various contexts and covered in various works (e.g., \cite{grzegorzewski2004measuring,grzegorzewski2006coefficient,grzegorzewski2009kendall,grzegorzewski2011spearman}), but it does not conform with the neutral treatment highlighted in this work. Therefore, it is not considered for the remainder of this paper.\bla \llV

\section{Handling incomplete rankings via correlation coefficients\llV}\label{Sec:IRviaCC}
To the best of our knowledge, there has not been a ranking correlation coefficient explicitly tailored to the space of non-strict incomplete rankings, $\Omega$, under a neutral treatment of incompleteness. Indeed, although \cite{emo02new} suggested that $\tx$ could fulfill this extended role, this assertion has not been formally proved nor empirically validated. Hence, the first of the ensuing subsections examines this hypothesis. Then, \S\ref{SS:tx_hat} introduces the ranking correlation coefficient $\txS$, along with the properties and axioms it satisfies. Finally, \S\ref{SS:pairings} establishes the equivalence of $\txS$ with the axiomatic distance $\ksN$ as well as the equivalence of $\tx$ with $\ksP$ when the input rankings lie in space $\Omega$. These connections are then leveraged into new exact optimization methodologies for solving NIRA in \S\ref{SS:optimization}. \llV

\subsection{Inadequacy of the Kendall Tau-Extended correlation coefficient}\label{SS:taux_inadequate}
This subsection provides cogent evidence that $\tx$ is not an adequate measure for quantifying and aggregating differences between incomplete rankings. Specifically, counter to what is claimed in \cite{emo02new}, employing $\tx$ produces incongruous and counterintuitive results when a judge's unranked objects should convey no preferential information. The veracity of these assertions is established via intuitive examples and the accompanying discussion.\llV

\begin{table}[H]\footnotesize\def\arraystretch{0.85}
\setlength{\tabcolsep}{3pt}
\centering
\begin{tabular}{cc!{\vrule width1pt}ccccccccccc!{\vrule width1pt}c}
           &    \multicolumn{ 12}{c}{{\bf \mmmH Input rankings}} & {\bf \footnotesize{$\tx-$Optima}}\mV \\
               && {\bf  $\av^1$} & {\bf  $\av^2$} & {\bf  $\av^3$} & {\bf  $\av^4$} & {\bf  $\av^5$} & {\bf  $\av^6$} & {\bf  $\av^7$} & {\bf  $\av^8$} & {\bf  $\av^9$} & {\bf  $\av^{10}$} & {\bf  $\av^{11}$} &  {\bf  $\rv^*_1$ \quad $\rv^*_2$}
    \\\hlineB{2.5}
\multirow{8}{*}{\rotatebox[origin=c]{90}{\bf \qquad \quad Objects}}&{\bf $v_1$} & 1 & 1 & $\nr$ & $\nr$ & $\nr$ & $\nr$ & $\nr$ & $\nr$ & $\nr$ & $\nr$ & 5  & 4 \quad 4  \\
&{\bf $v_2$} & 2  & 2 &  1 & 1 & $\nr$ & $\nr$ & $\nr$ & $\nr$ & $\nr$ & 1 & 4 & 5 \quad 5\\
&{\bf $v_3$} & $\nr$ & $\nr$ & 2 & 2 & 1 & 1 & 1 & $\nr$ & $\nr$ & $\nr$ & 3  & 2 \quad 1 \\
&{\bf $v_4$} & $\nr$ & $\nr$ & $\nr$ & $\nr$ & 2 & 2 & 2 & 1 & 1 & $\nr$ & 2 & 3 \quad 2  \\
&{\bf $v_5$} & $\nr$ & $\nr$ & $\nr$ & $\nr$ & $\nr$ & $\nr$ & $\nr$ & 2 & 2 & 2 & 1 & 1 \quad 3\\\hlineB{2.5}
&{\bf $\tx(\rv^*_1,\av^{k})$} & \small{$0.1$} & \small{$0.1$} & \small{$-0.1$} & \small{$-0.1$} & \small{$0.1$} & \small{$0.1$} & \small{$0.1$} & \small{$-0.1$} & \small{$-0.1$} & \small{$-0.1$} & \small{$0.6$} & \small{}\\
&{\bf $\tx(\rv^*_2,\av^{k})$} & \small{$0.1$} & \small{$0.1$} & \small{$-0.1$} & \small{$-0.1$} & \small{$0.1$} & \small{$0.1$} & \small{$0.1$} & \small{$0.1$} & \small{$0.1$} & \small{$-0.1$} & \small{$0.2$} & \small{}
\mmV 
\end{tabular}\llV\llV
\caption{Judge $\av^{11}$ wields outsize influence in the $\tx$ consensus rankings}\label{tab:txGral}\llV
\end{table}\llV

\bla Table \ref{tab:txGral} displays a simple instance consisting of incomplete rankings in space $\Omega$ over object set $V=\{v_1,\dots,v_5\}$; underneath the table are the underlying correlation values for each resulting optimal solution and each input ranking. Since, for every pair of objects $(v_i,v_{i+1})$ more judges prefer $v_i$ to $v_{i+1}$ than the reverse, it is reasonable from the majority's point of view to expect the identity permutation $\mathbf{\epsilon}_5:=(1,2,3,4,5)$ or a similar ranking as the consensus solution---10 of 11 people collectively express this belief. However, the $\tx$-consensus rankings are $\rv^*_1=(4,5,2,3,1)$ and $\rv^*_2=(4,5,1,2,3)$, which yield similarity values of $\tau(\rv^*_1,\mathbf{\epsilon}_5)=-0.2$ and $\tau(\rv^*_2,\mathbf{\epsilon}_5)=-0.6$, that is, they are highly dissimilar from the majority's preferences (by comparison, the all-ties ranking has a correlation with $\mathbf{\epsilon}_5$ of 0). Notice that  $\av^{10}$ and $\av^{11}$ are the only rankings that compare $v_2$ and $v_5$ and that, although $\av^{10}$ strictly prefers $v_2$ over $v_5$ and $\av^{11}$ strictly prefers $v_5$ over $v_2$, the optimal solutions only reflect the preferences of $\av^{11}$. More strikingly, $v_3$ is strictly preferred over $v_2$ in ${\rv}^*_1$ and ${\rv}^*_2$ even though two of the three judges who evaluate $(v_2, v_3)$ express the reverse preference. This suggests $\tx$ inadvertently imposes inequitable voting power in the aggregation process, specifically benefiting rankings with higher completeness ($\av^{11}$ in this example). Note that, in settings where the number of items to evaluate is unrestricted (e.g., product and travel ratings/reviews), awareness of this systematic biases could produce negative incentives such as spamming \citep{muk11det}.\bla

The inadequacy of $\tx$ to handle incomplete rankings can be discerned on a fundamental level from its inability to yield the extrema values 1 and $-1$ when an incomplete ranking is compared with itself and with its reverse ranking, respectively. In fact, the achievable correlation range shrinks as the number of ranked objects decreases. For instance, $\tx(\av,\bv)=-\frac{1}{3}$ when $\av=(1,2,\nr),\bv=(2,1,\nr)$, but $\tx(\av,\bv)=-\frac{1}{4}$ when $\av=(1,2,3,\nr),\bv=(3,2,1,\nr)$. This shrinking range translates into inequitability in the aggregation process. In Table \ref{tab:txGral}, for example, $\av^1$, $\av^2$, $\av^5$, $\av^6$ and $\av^7$ each obtain a $\tx$ correlation value of 0.1 with $\rv^*_1$ or $\rv^*_2$ even though each perfectly agrees over its ranked objects  
with both consensus rankings.

\bla A countering opinion to the preceding argument is that the full correlation range should not be achievable when comparing two incomplete rankings since it is possible these could increase or decrease in similarity once their preferences over all alternatives become known. However, this \textit{epistemic perspective} of missing data does not align with the neutral treatment adopted in this work, alternatively characterized as the \textit{ontic perspective} \citep{couso2014statistical}. The latter perspective more closely fits the purposes of this work since the objective is to piece together only the explicit preferences from each input so as to reach a solution that is equitable to each ranking input and representative of the collective preferences. Inferences of the individual's missing preferences is covered in literature outside of the current scope (e.g., see \citep{furnkranz2003pairwise,hullermeier2004comparison}). We remark that a similar perspective regarding missing data is adopted in \citep{coo07cre,dwo01ran} in the distance-based context (see \S \ref{SS:Distances} for details).\bla \llV

\subsection{Derivation of new correlation coefficient and its axiomatic foundation}\label{SS:tx_hat}
To quantify the similarity between non-strict incomplete rankings via correlation coefficients, a fundamental requirement is that the correlation between any pair of rankings $\av,\bv\in\Omega$ must lie within the interval $[-1, 1]$. The $-1$ and 1 values must be achieved whenever $\av$ and $\bv$ completely agree and completely disagree, respectively; otherwise, a value from the interior of the interval should be returned commensurate with the level of similarity. As explained in \S\ref{SS:taux_inadequate}, $\tx$ cannot fulfill these essential requirements. Hence, this subsection derives a new correlation coefficient that satisfies these properties as well as a set of metric-like axioms tailored to space $\Omega$. As a first step, we define a new ranking-matrix $[a_{ij}]$ representation for $\av\in\Omega$ as:\lV
\begin{flalign}
  a_{ij} & = \left\{\begin{array}{rl}
1 &\mmH \text{if } \mH a_i \le a_j,\\
-1 &\mmH \text{if } \mH a_i > a_j,\\
0 &\mmH \text{if } \mH i=j, \text{or } a_i=\bullet, \text{or } a_j=\bullet\\
\end{array}\right.\label{eqn:tauxN_sm1}
\end{flalign}
where $1\leq i,j\leq n$. This ranking-matrix can be obtained by extending Equation \eqref{eqn:taux_sm} to also assign $a_{ij}=0$ whenever object $i$ or $j$ (or both) is unranked in $\av$ and, thus, it is equivalent to the $\tx$ ranking-matrix when the input rankings are complete. This extension was cursorily proposed in \cite{emo02new}, although it was neither explicitly defined nor implemented therein. It is chosen as the basis of the new correlation coefficient also because its treatment of ties is equivalent to the Kemeny Snell ``half-flip'' metric, which assigns only half of a rank reversal between $\av$ and $\bv$ whenever one ties $(v_i,v_j)$ but the other professes a strict preference for $v_i$ over $v_j$, or vice versa.

As a second step, consider ranking-matrices $[a_{ij}]$ and $[b_{ij}]$ respectively defined according to Equation \eqref{eqn:tauxN_sm1} and their associated \textit{matrix inner product}:\llV
\[\sum_{i=1}^n\sum_{j=1}^{n}a_{ij}b_{ij}.\llV\]
When $\av$ and $\bv$ rank every object, the number of non-zeros in each ranking-matrix and the maximum matrix inner product are both equal to $n(n\mi1)$. The reasons are that the ranking-matrix diagonal elements are all 0 and that $a_{ij}b_{ij}=1$ for all $i\ne j$ when $b_{ij}=a_{ij}$. It is also straightforward to discern that a minimum matrix inner product of $\mi n(n\mi1)$ can be achieved only if $\av$ does not contain ties and $b_{ij}=\mi a_{ij}$ for all $i\neq j$.

When $\av$ or $\bv$ does not rank every object, for each $v_i$ such that either $a_i=\nr$ or $b_i=\nr$, the $i$th ranking-matrix row and column are set to zero, thereby decreasing the maximum and increasing the minimum matrix inner products by $2(n\mi1)$. Put otherwise, such a matrix inner product may be calculated as if the $i$th row and column of both ranking-matrices do not exist. Hence, the maximum and minimum inner products of $[a_{ij}]$ and $[b_{ij}]$ are reduced to $\bar n(\bar n-1)$ and $\mi\bar n(\bar n-1)$, respectively, where $\bar n=\abs{\AB}$. Accordingly, a new correlation function can be  derived to achieve the full expected correlation interval $[\mi1,1]$. It is named the \textit{scaled Kendall tau-extended correlation coefficient}, written succinctly as $\txS$, and is defined as:
\begin{flalign}
    \txS(\av,\bv)&= \frac{\sum_{i=1}^{n}\sum_{j=1}^{n}a_{ij}b_{ij}}{\bar n(\bar n-1)},\label{eqn:ntaux}
\end{flalign}
which may be rewritten in terms of $\tx$ via the equation: \llV\lV
\begin{flalign}
    \txS(\av,\bv)&= \frac{n(n-1)}{\bar n(\bar n-1)}\tx(\av,\bv), \label{eqn:ntaux_other}
\end{flalign}
assuming the underlying ranking-matrix of $\tx$ is given by Equation \eqref{eqn:tauxN_sm1}. This alternative expression emphasizes that, by scaling $\tx(\av,\bv)$ by the factor $\frac{n(n-1)}{\bar n(\bar n-1)}\geq1$, $\txS$ removes the impact of irrelevant pairwise preference comparisons---the pairs of objects unranked by $\av$ or $\bv$---from their correlation. As a result, the extrema correlation values $-1$ and 1 can be achieved when comparing two appropriate non-strict incomplete rankings. 

Clearly, when $\AB=V$, the Equation \eqref{eqn:ntaux_other} scaling factor equals 1, meaning $\txS$ is equivalent to $\tx$ in space $\Omega_C$; 
Furthermore, $\txS$  is equivalent to $\t$ in space $\Omega_C\cap\Omega_S$ due to the equivalence between $\tx$ and $\t$ in said space \citep{emo02new}. Hence, $\txS$ possesses the same advantages as $\tx$ and $\t$ when the rankings are restricted to spaces $\Omega_C$ and $\Omega_C\cap\Omega_S$, respectively.  The remainder of this subsection will provide a theoretical basis for why $\txS$ is uniquely suited to deal with the broader space of non-strict incomplete rankings $\Omega$; additional practical reasons are given by the empirical results obtained in \S\ref{SS:Experiments_1}--\S\ref{SS:Experiments_3}.

\bla This paragraph presents the set of intuitive metric-like axioms that $\txS$ satisfies; formal proofs are in Appendix \ref{App:axioms_coefficient}. To present the $\txS$ axiomatic foundation, one concept must be introduced. Namely, $\bv$ is said to be \textit{between} $\av$ and $\cv$ if, for each  $(v_i,v_j)$, the preference judgment of $\bv$ either (i) agrees with $\av$ or (ii) agrees with $\cv$ or (iii) $\av$ prefers $v_i$, $\cv$ prefers $v_j$, and $\bv$ ties them. 

\begin{axiom}[Relevance] The correlation discounts the unevaluated objects:\llV\lV
$$\txS(\av,\bv)=\txS(\ap,\bp).$$\end{axiom}\lllV

\begin{axiom}[Commutativity] The correlation value is independent of the order in which  $\av$ and $\bv$ are compared: \llV\lV
\[\txS(\av,\bv)=\txS(\bv,\av).\]\end{axiom} \lllV

\begin{axiom}[Neutrality] The correlation value is independent of the particular labeling of the objects:

If $\av'=\pi(\av)$ and $\bv'=\pi(\bv)$, then $\txS(\av,\bv)=\txS(\av',\bv')$, where  $\pi:=\{1,2,...,n\} \rightarrow \{1,2,...,n\}$ is a permutation function.\end{axiom}\lllV

\begin{axiom}[Reduction] 
If $\av$ and $\bv$ agree except for a set $V'\subseteq V$, then $\txS(\av,\bv)$ may be computed by focusing only on the objects in $V'$:  \llV
\begin{equation*}
    \txS(\av, \bv) = 1 + 2\txS(\av|_{V'}, \bv|_{V'}).
\end{equation*} 
\end{axiom} \lllV

\begin{axiom}[Relaxed Triangle Inequality] Relationship among the three possible paired comparisons from three incomplete rankings: \llV\lV
\begin{equation*}
    \txS(\av\pte,\bv\pte)+\txS(\bv\pte,\cv\pte) \leq \txS(\av\pte,\cv\pte)+1;\llV
\end{equation*} and equality holds if and only if $\bv\pte$ is between the other two projected rankings; here, $V_{\av,\bv,\cv} := \sA\inter \sB\inter\sC$ for concise representation.  \end{axiom}\lllV

\begin{axiom}[Scaling] The correlation range is between -1 and 1, inclusively:
\begin{equation*}
-1\leq \txS(\av,\bv)\leq1;\llV 
\end{equation*}
with $\txS(\av,\bv)=1$ iff $\ap=\bp$ and $\txS(\av,\bv)=-1$ iff  $\bp$ is the reverse ranking of $\ap$ (the latter must be a linear ordering). \end{axiom}\llV

Axiom 6 ensures equitability in the aggregation process regardless of how many objects are ranked by each judge. It also brings the pragmatic benefit of eliminating the unenforceable/unrealistic requirement of having to allocate an equal number of objects for each judge to evaluate, which may be difficult to enforce due to differing expertise, disagreeing schedules, unplanned exemptions, etc. \citep{hoc10all}. Indeed, it is advisable to avoid assigning fewer subjective evaluation tasks to mitigate cognitive errors \citep{bas15cho,saa03mag}

\bla As suggested by Axiom 1, the $\txS$ similarity between two incomplete rankings can be equivalently calculated by simply dropping the alternatives unranked by either ranking (i.e., by projecting them to the subset of objects evaluated by both). While this may seem to remove the incomplete data from the researcher's view when comparing two incomplete rankings, we emphasize that the consensus ranking problem (see \eqref{eqn:c-consensus} or \eqref{eqn:d-consensus}) involves accruing the comparisons between the candidate solution (always a complete ranking) and each input ranking (which may be incomplete or complete). In this context, Axiom 1 ensures that each input incomplete ranking influences only the consensus ranking elements corresponding to its ranked objects.\bla \llV

\subsection{Key pairings between distances and correlation coefficients}\label{SS:pairings}
This subsection establishes a formal connection between $\txS$ and $\ksN$ as well as between another key axiomatic-distance and correlation-coefficient pairing in space $\Omega$. Together these results fill a significant gap in the literature because although \cite{emo02new} made a connection between distance and correlation-based methods for complete rankings (see Equation \eqref{eqn:taux_to_ks}), they conjectured that a parallel connection could not be established for incomplete rankings. \bla Proofs to all related theorems and corollaries are in \S\ref{App:proof_thm_cor}\bla. 

\begin{theorem}[Linear transformation between $\txS$ and $\ksN$]\label{thm:txS_to_ksN}
Let $\av$ and $\bv$ be arbitrary rankings over $n=|V|$ objects drawn from the space of non-strict incomplete rankings, $\Omega$. Then, the $\txS$ correlation coefficient and the $\ksN$ distance are connected through the equation:
\begin{equation}\label{eqn:txS_to_ksN}
  \ksN(\av,\bv)=\frac{1}{2}-\frac{1}{2}\txS(\av,\bv).\llV
\end{equation}
\end{theorem}\llV

\begin{theorem}[Linear transformation between $\tx$ and $\ksP$]\label{thm:tx_to_ksP}
Let $\av$ and $\bv$ be arbitrary rankings of $n=|V|$ objects from space $\Omega$. Then, the $\tx$ correlation coefficient and the $\ksP$ distance are connected through the equation:
\begin{equation}\label{eqn:tx_to_ksP}
  \ksP(\av,\bv)=\frac{\bar n(\bar n\mi1)}{4}-\frac{n(n-1)}{4}\tx(\av,\bv), \llV
\end{equation}
where $\bar n=|\AB|$ (i.e., the number of objects ranked by both $\av$ and $\bv$).
\end{theorem}\llV

The following two corollaries are a direct result of these connections.

\begin{corollary}\label{cor:unbiasedEquiv}
The NIRA optimization problems typified by $\txS$ and $\ksN$ are equivalent and, thus, provide identical consensus rankings. Similarly, the NIRA optimization problems typified by $\tx$ and $\ksP$ are equivalent.
\end{corollary}\llV

\bla \begin{corollary}\label{cor:nphard}
The correlation-based NIRA is $\cal{NP}$-hard.
\end{corollary}
\bla

The distance-correlation pairings provide mutal support for the usefulness of the respective measures. In particular, $\tau$, $\tx$, and $\txS$ are strengthened by the robust properties and social choice foundations of the Kemeny aggregation framework (see \citep{bra16han,kem62pre,you88con,you78con}). Meanwhile, $\ks$, $\ksP$, and $\ksN$ benefit from the computational advantages engendered by the correlation-based framework, which include a linearized NIRA objective function. These advantages are bolstered by the optimization methodologies developed in the next subsection. \llV

\subsection{Exact optimization methodologies}\label{SS:optimization}
\citeauthor{emo02new} \cite{emo02new} devised a branch and bound algorithm for solving the ranking aggregation problem that relies on a succinct function of cumulative agreement between the set of input rankings $\{\av^k\}_{k=1}^K$ and an iteratively evolving candidate-solution vector $\rv\in\Omega_C$. In this respect $\tx$ offers a significant advantage over its distance counterpart $\ksP$ since: \llV
\begin{align}\label{eqn:txCI}
\sum_{k=1}^{K}\tx(\rv,\av^k)
  &=\sum_{k=1}^{K}\sum_{i=1}^{n}\sum_{j=1}^{n}\frac{a^k_{ij}r_{ij}}{n(n\mi1)}=
  \frac{1}{n(n\mi1)}\sum_{i=1}^{n}\sum_{j=1}^{n}A_{ij}r_{ij},
\end{align}
where $[a^k_{ij}]$ and $[r_{ij}]$ represent the ranking-matrices of $\av^k$ and $\rv$, respectively; and where $[A_{ij}]=\sum_{k=1}^{K}a^k_{ij}$ is defined as the \textit{combined ranking-matrix} (CR). Once the CR is computed, the number of matrix inner products required to calculate the objective function value of a candidate solution is reduced from $K$ to one. Conversely, the cumulative distance function for $\ksP$ (or for $\ks$ or $\ksN$) does not yield as wieldy of an expression due to the presence of nonlinear terms. CR's computational advantage thus enables a form of sensitivity analysis used in the B\&B algorithm for determining the increase/decrease in cumulative agreement that results when the preference or ordinal relationships of a few objects in a candidate ranking are altered. 

Since each correlation coefficient term $\txS(\rv, \av^k)$ in the $\txS$ NIRA objective function has a different denominator---based on the number of objects ranked by $\av^k$---CR and the \citeauthor{emo02new} B\&B algorithm are inapplicable for $\txS$. To fix this issue, the following theorem introduces a corresponding succinct function of cumulative agreement (see \ref{App:proof_thm_cor} for its proof). 

\begin{theorem}[Succinct function of cumulative agreement for $\txS$]\label{thm:txS_cumu}
Let $\rv\in\Omega_C$, $\av^k\in\Omega$, and $\bar n^k = |V_{\av^k}|$ (the number of objects ranked by $\av^k$), for $k=1,\dots,K$. Then, the $\txS$ cumulative correlation between $\rv$ and $\{\av^k\}_{k=1}^K$ can be computed according to the function:\llV
\begin{equation}\label{eqn:txSCI}
      \sum_{k=1}^{K}\txS(\rv, \av^k)=
\sum_{i=1}^{n}\sum_{j=1}^{n}\hat A_{ij}r_{ij},\llV\lV
\end{equation}
where $[\hat A_{ij}]:=\sum_{k=1}^{K}\frac{a^k_{ij}}{\bar n^k(\bar n^k-1)}$ is the \textit{scaled combined ranking-matrix (SCR)}.\llV
\end{theorem}
Based on this expression, it is possible to extend the specialized branch and bound algorithm from \citep{emo02new} to find the comprehensive solution to NIRA. The resulting B\&B algorithm returns the complete NIRA optimality set through an efficient exploration of the solution space, whereby unpromising branches (i.e., ordinal combinations) are pruned to avoid enumeration. Following the calculations of $[\hat A_{ij}]$ and $[A_{ij}]$, the steps of the B\&B algorithms for $\txS$ and $\tx$ are identical. \bla For the reader's convenience, a flowchart and detailed description of the NIRA B\&B algorithm are given in Appendix \ref{App:b&b}.\bla 

\bla The combination of Theorems \ref{thm:txS_to_ksN}-\ref{thm:txS_cumu} and Corollary \ref{cor:unbiasedEquiv} enable the formulation of the first exact integer programming models for NIRA. In particular, \citeauthor{yoo2018newinteger} \citep{yoo2018newinteger} recently derived a set of linear constraints that imposes logical conditions for a matrix of decision variables $R\in\{\mi1,0,1\}^{n\times n}$ to induce all  
ranking-matrices of non-strict complete rankings (see Equation \eqref{eqn:taux_sm}). The feasible solutions to these constraints reflect all possibilities for the solution vector $\rv\in\Omega_C$ and, hence, they can be paired with Expression \eqref{eqn:txCI} or \eqref{eqn:txSCI} to solve the NIRA via $\tx$ (equivalent to $\ksP$) or via $\txS$  (equivalent to $\ksN$) respectively. While these formulations are faster to solve than B\&B, they are guaranteed to return only one optimal solution. As such, they are not suited for the computational studies herein conducted. For the reader's convenience, the NIRA formulation for $\txS$ is given in Appendix \ref{App:IP}.\bla \llV

\section{Computational studies on the correlation-based NIRA\llV}\label{Sec:Algorithms}
This section seeks to assess the comparative abilities of $\tx$ and $\txS$ to be \textit{decisive} and \textit{electorally fair}. \bla The decisiveness property---characterized by the propensity to yield unique or few alternative optima---is pivotal because many competitive group decision-making scenarios entail the allocation of a limited budget among a subset of participants proportional to the consensus ranking solution. In these situations, it is prudent to obtain the full set of alternative optimal solutions when the consensus ranking is not unique. Reporting only one consensus in these cases would engender unfair and/or arbitrary outcomes. However, by Corollary \ref{cor:nphard}, obtaining even just one consensus ranking via correlation-based methods (or the equivalent axiomatic distance-based methods \citep{bar89vot}) is an $\cal{NP}$-hard problem. What is more, since $|\Omega|\approx.5[(1.4)^{n+1}n!]\gg n!$ \citep{gro62pre}, enumerating the full solution space is impractical except for a very small number of objects---for $n=10$ and $n=20$, there are approximately $1\times10^9$ and $2.5\times 10^{21}$ solutions, respectively. The electoral fairness property is a central democratic tenet in social choice. Because each judge wishes the consensus ranking to be as close as possible to his/her preferences \citep{contucci2016egalitarianism}, ranking aggregation methods should maximize the agreement of the solution with as many of the inputs as possible to be equitable. This property is assessed by measuring the similarity between the consensus ranking to an underlying ground truth shared by a majority\bla. To evaluate both properties empirically, the ensuing subsection develops a probabilistic approach for generating instances from space $\Omega$. \llV

\subsection{Generation of representative instances from space $\Omega$}\label{SS:Sampling}
In the computational study carried out in \citep{mor16axi}, $\ksN$ outperformed $\ksP$ in yielding fewer alternative optima when solving instances with ``predictable consensus rankings'' of the distance-based NIRA. Although these results seem to support the premise that $\txS$ is better suited than $\tx$ to solve the correlation-based NIRA (due to Corollary \ref{cor:unbiasedEquiv}), their scope is limited based on the restrictive types of instances therein considered. Specifically, since enumeration was employed to solve the problems exactly, each tested instance consisted of at most 15 non-strict incomplete rankings of dimension $n=7$. These were generated from three simplistic templates. The first two initialize every input to a reference ranking $\avt\in\Omega_C$---the all-ties ranking and the identity permutation, $\mathbf{\epsilon}_n$, respectively---and a third differs slightly from the second in that it initializes a minority of inputs to the reverse ranking of $\mathbf{\epsilon}_n$. For all three templates, incompleteness is inserted to a random number and selection of objects within each initialized ranking. The incorporated level of individual agreement/disagreement of these instances is extreme and, thus, uncharacteristic of most group decision-making settings.

To generate test instances that are representative of different group decision-making scenarios, we devise a sampling approach based on Mallow's $\phi$-model of ranking data \cite{mal57non}, which is tailored to distance-based methods \citep{mar96ana}. The standard $\phi$-model  is parameterized by a reference or ``ground truth'' ranking $\avt\in\Omega_C$ and dispersion $\phi\in(0,1]$ which, in conjunction with the $\ks$ distance, quantify the probability of observing a ranking $\av\in\Omega_C$ as:\llV
\[P(\av)=P(\av|\avt,\phi)=\frac{1}{Z}\phi^{\ks(\av,\scriptsize\avt)},\llV\] 
where $Z=\Sigma_{\rv\in\Omega_C}\phi^{\ks(\rv,\scriptsize\avt)}=(1)\times(1+\phi)\times(1+\phi+\phi^2)\times\dots\times(1+\dots+\phi^{n-1})$ is the normalization constant. Since setting $\phi$ to 1 yields the uniform distribution over space $\Omega_C$ and setting it nearer to 0 centers the distribution mass closer to $\avt$, the dispersion parameter effectively characterizes the proximity of $\av$ to $\avt$ \citep{lu14eff}. \bla Note that fixing the same dispersion value to generate all input rankings of an instance will yield rankings that have approximately the same degree of similarity with $\avt$. Instances with lower (higher) $\phi$ values will yield instances that are more (less) cohesive collectively. Put otherwise, higher $\phi$ values will generate non-trivial instances with higher noise levels.\bla

 Since sampling directly from the $\phi$-distribution can be very inefficient, we adapt the repeated insertion model (RIM) developed in \cite{doi04rep} for the efficient unconditional sampling of non-strict incomplete rankings. On a related note, we do not use the Generalized RIM model \cite{lu14eff} since it is a conditional approach that relies on assumptions that are more appropriate for implicitly gathered choice data. In RIM, a set of specified insertion probabilities $p_{ij}$ for $1\le j\le i\le n$ is used to efficiently construct a random ranking $\av$ from the reference ranking $\avt\in\Omega_C$. Specifically, assuming objects $\avt^{\mi1}(1),\dots, \avt^{\mi1}(i\mi1)$ have been assigned certain ranking positions within $\av$, where $\avt^{\mi1}(j)$ is a function that returns the $j$th-highest ranked object in $\avt$ (see Appendix \ref{App:b&b} for more details). Object $\avt^{\mi1}(i)$ is then inserted at rank $j\le i$ (i.e., assigned position $j$ in $\av$) with probability $p_{ij}$, for $i=1,\dots,n$. The choice of insertion probabilities $p_{ij}$=$\phi^{i-j}/(1+\phi+\dots+\phi^{i-1})$ guarantees that $\sum^i_{j=1}p_{ij}=1$ for all $i$, and it induces the standard Mallows $\phi$-distribution \citep{doi04rep}. To generate incomplete rankings by a similar procedure, we propose two natural extensions for RIM. With this intent in mind, assume that the object subset to be ranked by $\av$, $V_{\av}$, is known a priori. Pseudocodes of these insertion models, abbreviated as RIME1 and RIME2, are shown in Algorithms \ref{RIME1_pseudo} and \ref{RIME2_pseudo}. \llV
 
\begin{algorithm}
\footnotesize
    \caption{Repeated Insertion Model Extension \#1 (RIME1)}
	\textbf{Input}: Object set for $\av$: $V_{\av}$, projected reference object ordering: $\avt^{-1}|_{V_{\av}}$, dispersion: $\phi$\\
	\textbf{Output}: Generated incomplete ranking subsample

	\begin{algorithmic}[1]
		\FOR{$i=1, 2, ..., |V_a|$}
        \FOR{$j=1, 2, ..., i$}
			\STATE $a^{-1}_j \gets {\underbar{$a$}^{-1}_i}|_{V_{\av}}$ with probability: $p_{ij}=\phi^{i-j}/(1+\phi+\dots+\phi^{i-1})$
        \ENDFOR
        \ENDFOR
        \FOR{$i=1, 2, ..., |V|$}
            \IF{$v_{i} \in V_{\av}$}
		    	\STATE $a_i \gets$ rank position of $v_i$ in $\av^{-1}$
		    \ELSE
		        \STATE $a_i \gets \nr$
	    	\ENDIF
        \ENDFOR
	\end{algorithmic}
	\label{RIME1_pseudo}
\end{algorithm}\lllV
\begin{algorithm}[h!]
\footnotesize
	\caption{Repeated Insertion Model Extension \#2 (RIME2)}
	\textbf{Input}: Object set for $\av$: $V_{\av}$, reference object ordering: $\avt^{-1}$, dispersion: $\phi$\\
	\textbf{Output}: Generated incomplete ranking subsample
	
	\begin{algorithmic}[1]
		\FOR{$i=1, 2, ..., |V|$}
        \FOR{$j=1, 2, ..., i$}
			\STATE $a^{-1}_{j} \gets \underbar{$a$}^{-1}_i$ with probability: $p_{ij}=\phi^{i-j}/(1+\phi+\dots+\phi^{i-1})$
        \ENDFOR
        \ENDFOR
        \FOR{$i=1, 2, ..., |V|$}
            \IF{$v_{i} \in V_{\av}$}
		    	\STATE $a_i \gets$ rank position of $v_i$ in $\av^{-1}$
		    \ELSE
		        \STATE $a_i \gets \nr$
	    	\ENDIF
        \ENDFOR
	\end{algorithmic}
	\label{RIME2_pseudo}
\end{algorithm}\llV

Explored in greater detail, RIME1 applies RIM to generate a ranking over $V_{\av}$ belonging to the Mallows $\phi$-distribution parameterized by pair ($\avt|_{V_{\av}},\phi)$, which is then expanded to $V$ by setting the positions of unranked objects to $\bullet$. RIME2 applies RIM to generate a complete ranking over $V$ belonging to the Mallows $\phi$-distribution parameterized by pair ($\avt,\phi)$, which is then converted into an incomplete ranking by keeping the numerical values of only the objects in $V_{\av}$ (i.e., replacing the positions of all other objects with $\bullet$). To explain the differences between RIME1 and RIME2 more intuitively, assume that in $\avt$, object $v_i\in V$ is strictly preferred over all objects in a nonempty subset $V'\subset V$, all of which are in turn strictly preferred over object $v_j\in V$. That is, in the ground truth, $v_i$ is strictly preferred over $v_j$ and $V'$ are objects with intermediary ordinal positions. The core distinction between the two insertion models is that, as the subset of unranked objects (i.e., $V'$) for a given judge increases in size, RIME2 proportionally decreases the probability that the reference ordinal positions for $v_i$ and $v_j$ will be reversed in $\av$. Conversely, RIME1 determines the probability of this event as if $v_i$ and $v_j$ have consecutive ordinal positions in $\avt$, thus ignoring the reference positions of unranked objects. In other words, unlike RIME1, RIME2 implicitly incorporates a relative ``intensity of preference'' \citep{coo06dis,coo85ord} between $v_i$ and $v_j$ that reflects the intermediary ordinal positions $\avt$ assigns to $V'$ even though this subset is not explicitly considered by $\av$.  Hence, the two insertion models generate non-strict incomplete ranking instances from opposing viewpoints regarding preferences over unranked objects. Although these models do not capture every possible assumption for such preferences, they can be utilized to generate nontrivial random instances with controllable degrees of collective similarity from which general  conclusions about the behavior of individual ranking measures can be drawn. It is important to mention that alternative sampling approaches and specialized distributions of ranking data could also be extended to generate random instances of incomplete rankings (e.g., the Plackett-Luce model \citep{luc59ind,pla75ana}). A thorough comparison of these alternatives and a formal analysis of the statistical distributions induced by RIME1 and RIME2 are left for future work. \llV

\subsection{Assessing decisiveness}\label{SS:Experiments_1}
This subsection concentrates on assessing the desired practical property of decisiveness. The property is especially relevant owing to the fact that ranking aggregation algorithms are typically designed to yield only one of an instance's possible multiple rankings, for the sake of efficiency. (For instance, \cite{mor16axi} devised an exact algorithm for $\ksN$ based on the implicit hitting set approach \citep{mor13imp} that returns exactly one optimal solution in $\Omega_C\cap\Omega_S$). Thus, even though obtaining more than one optimal solution or certifying unique optimality may be infeasible in practice, it would be reassuring for decision makers to know a priori that certain robust measures tend to return relatively fewer alternative optimal solutions than others. Accordingly, the decisiveness of $\tx$ and $\txS$ is assessed by how each measure curtails the growth in the number of alternative optimal solutions as data becomes noisier. To this end, this section performs experiments on random instances generated with RIME1 and RIME2 and characterized according to the four simple real-world group decision-making scenarios described in Table \ref{table:scenarios}.  

\begin{table*}[ht]
\tiny \centering
	\caption{Characterization of rankings generated from one ground truth and one dispersion}\llV
	\label{table:scenarios}
	\renewcommand{\arraystretch}{1.5}
	\begin{tabular}{|p{1.75cm}|p{4.4cm}|p{4.6cm}|l}
		\cline{1-3}
		 $\phi$ (Dispersion) & \bf General group characterization & \bf Example scenarios &
		 \\
		\cline{1-3}
		$\phi \in (0, 0.25]$ & \textbf{Strong collective similarity} \newline ``Subject-matter experts'' & $\bullet$ Federal grant proposal reviews \newline $\bullet$
		Olympic events with a style component  \newline $\bullet$
		Standardized test essay grading & \multirow{2}{*}{\tikz[remember picture,overlay] \coordinate (a);\hspace*{2ex}\shortstack[l]{\textbf{Low}\\\rotatebox{270}{\qquad \textbf{subjectivity}}}}\\ \cline{1-3}
		$\phi \in (0.25, 0.50]$ & \textbf{Weak collective similarity} \newline ``Seasoned body of objective observers'' & $\bullet$ University rankings \newline $\bullet$ Paid movie critiques \newline $\bullet$ Official World Cup ranking forecasts &  \\ \cline{1-3}
		$\phi \in (0.50, 0.75]$ & \textbf{Weak collective dissimilarity} \newline ``Public with background information" & $\bullet$ Unpaid movie recommendations  \newline $\bullet$ Unsponsored top travel lists \newline $\bullet$ Car brand preferences  &\\ \cline{1-3}
		$\phi \in (0.75, 1]$ & \textbf{Strong collective dissimilarity} \newline ``Jumble of heterogeneous opinions" & $\bullet$ Favorite colors \newline $\bullet$ Luckiest numbers  & \multirow{2}{*}[.5em]{\tikz[remember picture,overlay] \coordinate (c);\hspace*{2ex}\textbf{High}} \\ \cline{1-3} 
	\end{tabular}
	\tikz[remember picture,overlay] \draw[<->, very thick] (a)--(c);
\end{table*}

For the parameter configurations detailed below, the number of alternative optima returned by each correlation coefficient is individually recorded for 10 corresponding instances and summarized via average (AVG) and standard deviation (SD) values (\bla graphs display AVG and AVG$\pm$SD values via error bars\bla). Test problems are solved exactly via B\&B (see Appendix \ref{App:b&b}) until the full solution space is fathomed. Since B\&B follows a nearly identical logic when executed with $\tx$ or $\txS$ and run-time decisions are instance-specific, differences in solution times were insufficiently favorable in either direction and are unrecorded. For a more efficient version of B\&B, which has been shown to provide good quality solutions, but which does not guarantee global optimality nor return alternative optima, see \cite{amo16acc}.

All experiments were performed on machines equipped with 22GB of RAM memory shared by two 2.8 GHz quad core Intel Xeon 5560 processors; code was written in Python.

\begin{figure*}[ht!]
    \begin{subfigure}{0.95\textwidth}
    \centering
		\includegraphics[width=0.2in]{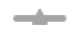}
		$\tx$
		\includegraphics[width=0.2in]{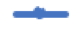}
		$\txS$  
    \end{subfigure}
    \newline
	\begin{subfigure}{0.49\textwidth}
		\centering
		\includegraphics[width=2.4in,height=1.2in]{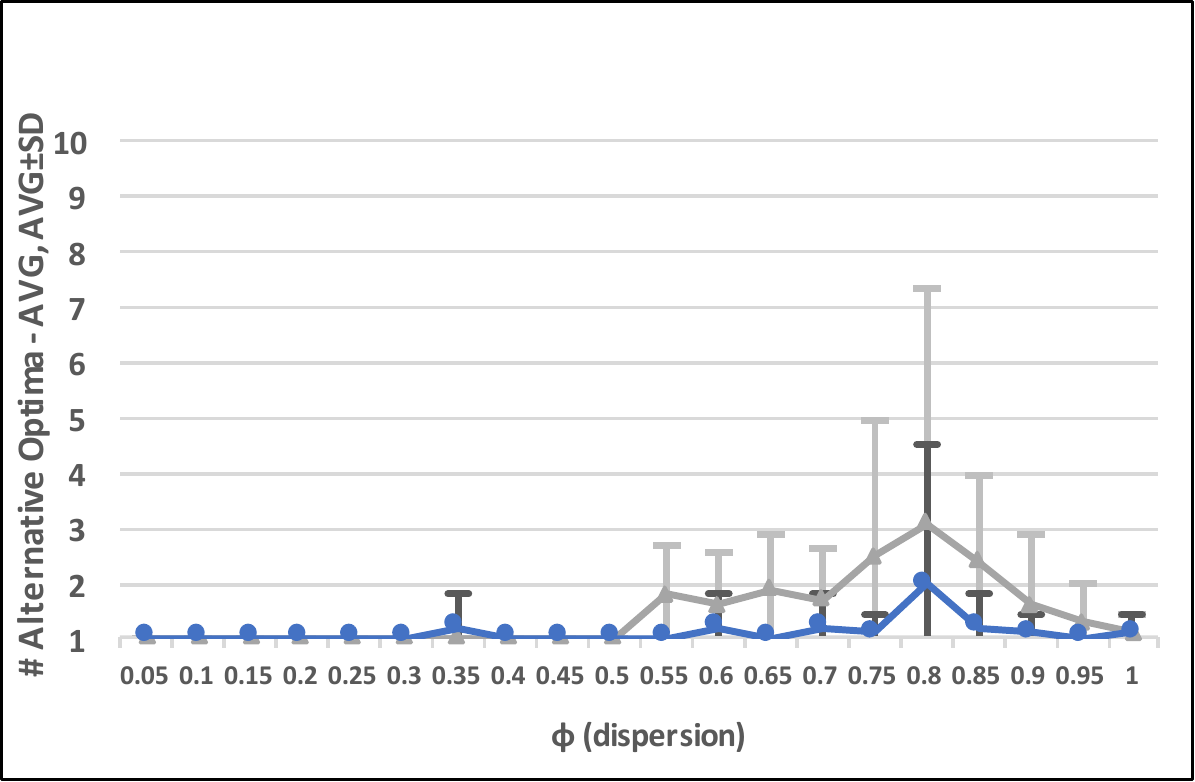}
		\caption{ RIME1 Instances with $|V|=8$}
		\label{fig:rime1_8}
	\end{subfigure}
    \begin{subfigure}{0.49\textwidth}
		\centering
		\includegraphics[width=2.4in,height=1.2in]{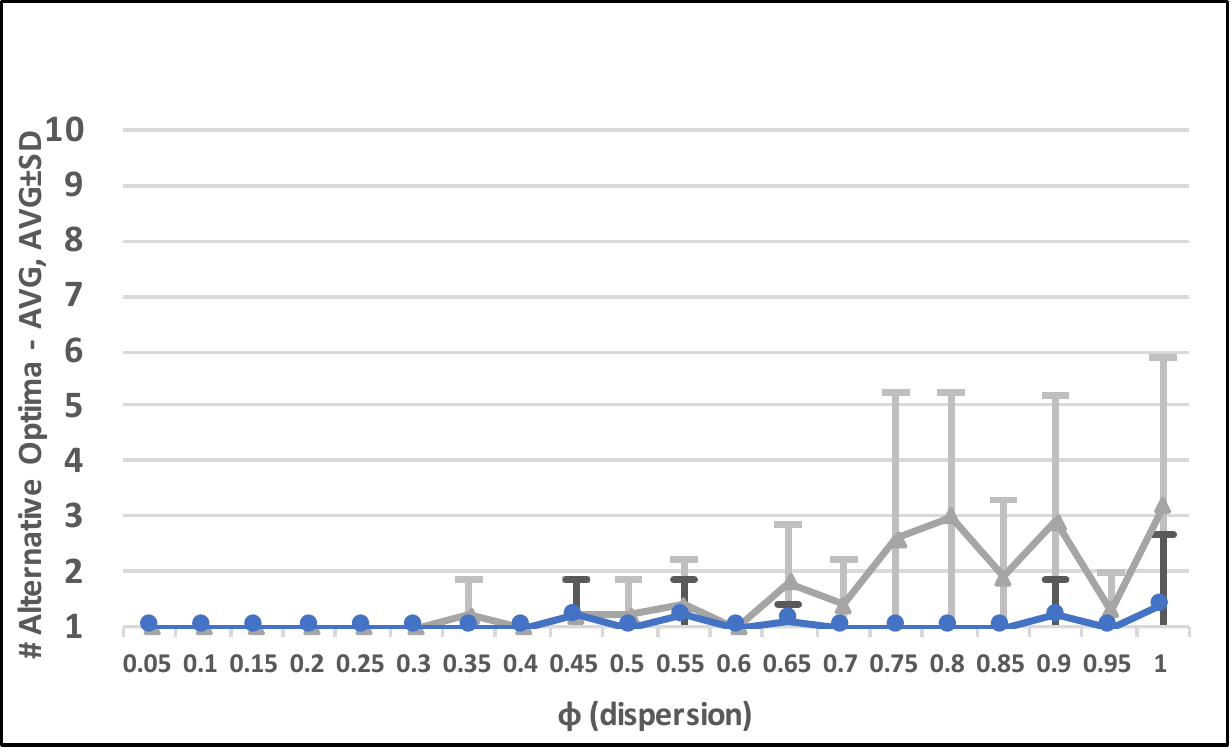}
		\caption{RIME2 Instances with $|V|=8$}
		\label{fig:rime2_8}
	\end{subfigure}
	\begin{subfigure}{0.49\textwidth}
		\centering
		\includegraphics[width=2.4in,height=1.2in]{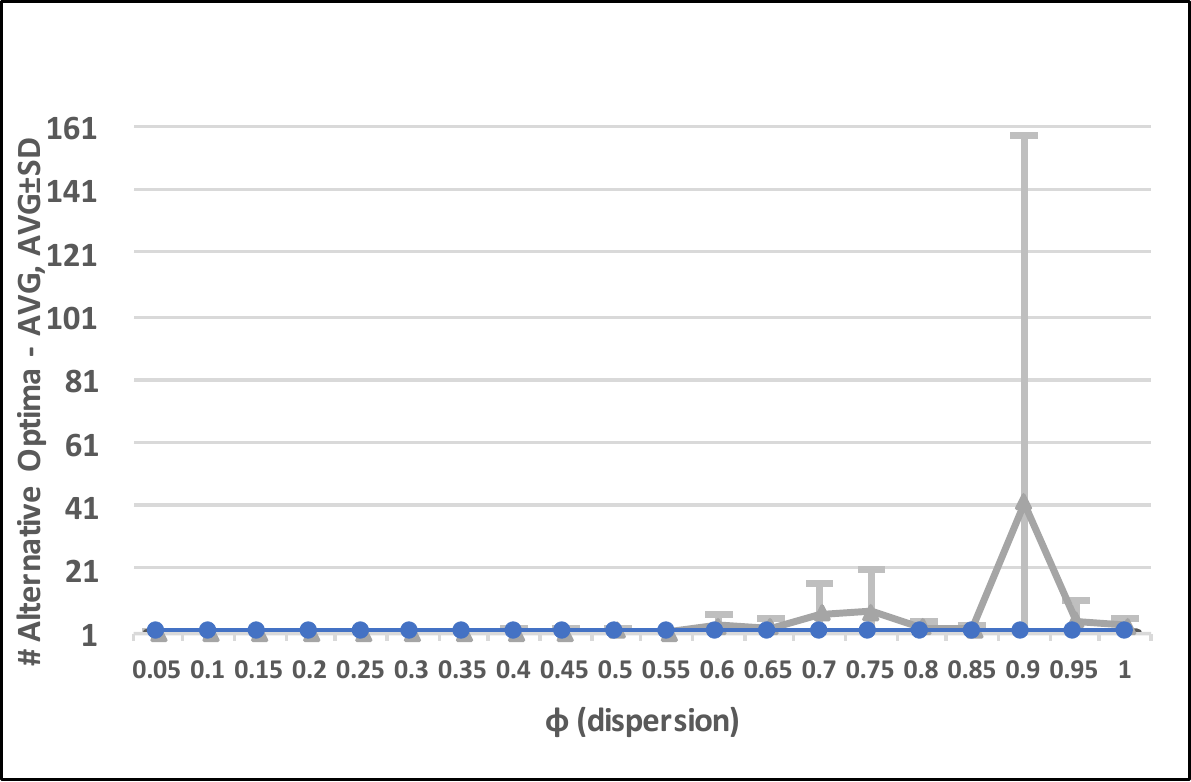}
		\caption{RIME1 Instances with $|V|=12$}
		\label{fig:rime1_12}
	\end{subfigure}
    \begin{subfigure}{0.49\textwidth}
		\centering
		\includegraphics[width=2.4in,height=1.2in]{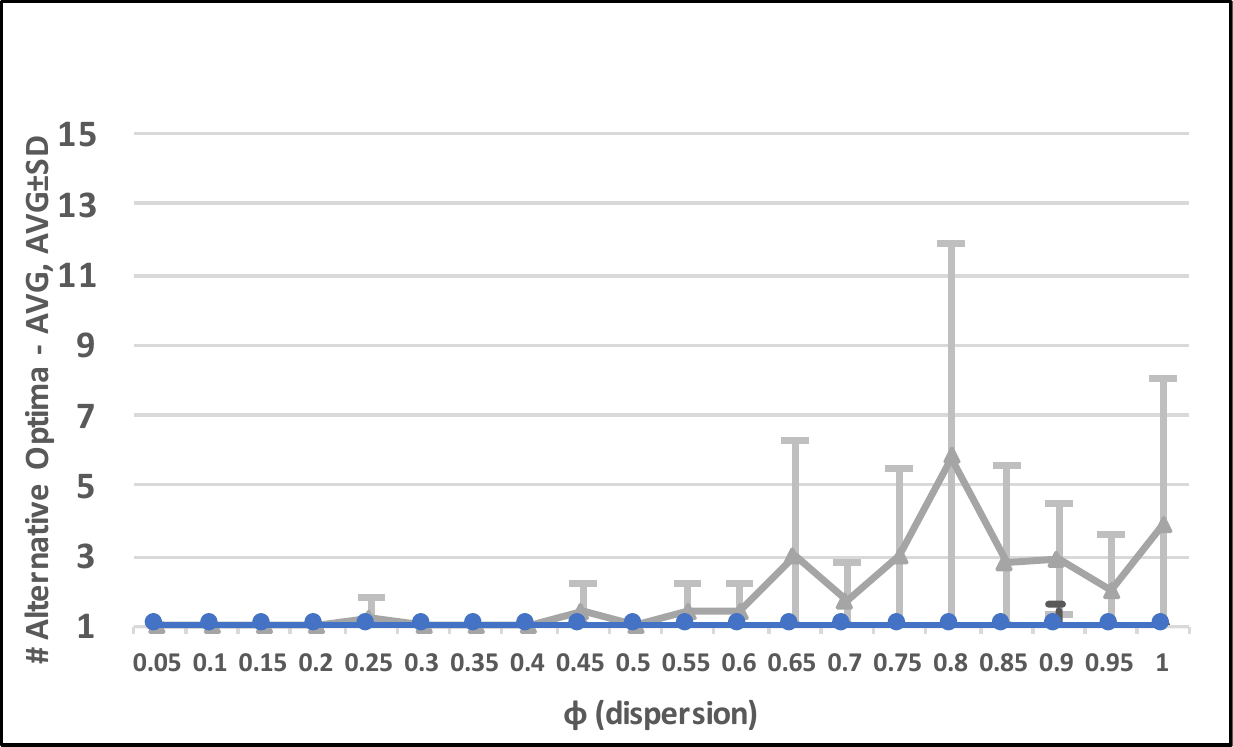}
		\caption{RIME2 Instances with $|V|=12$}
		\label{fig:rime2_12}
	\end{subfigure}
	\begin{subfigure}{0.49\textwidth}
		\centering
		\includegraphics[width=2.4in,height=1.2in]{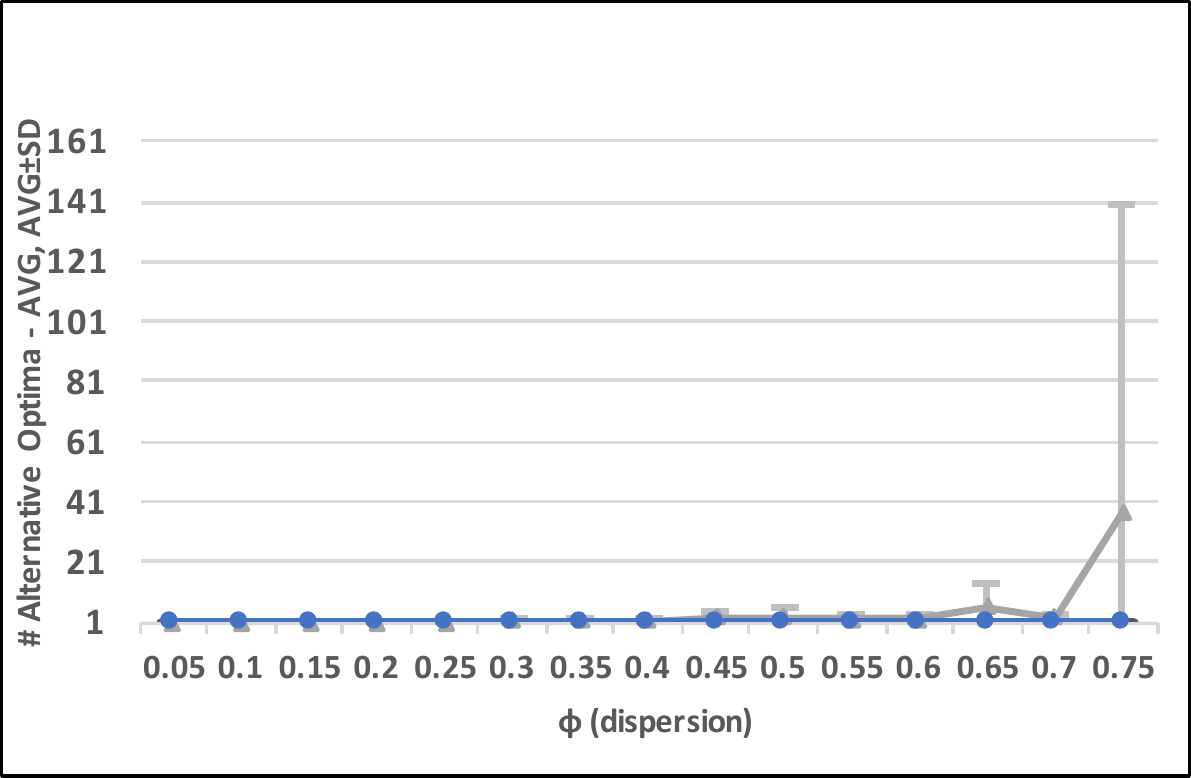}
		\caption{RIME1 Instances with $|V|=16$}
		\label{fig:rime1_16}
	\end{subfigure}
	\begin{subfigure}{0.49\textwidth}
		\centering
		\includegraphics[width=2.4in,height=1.2in]{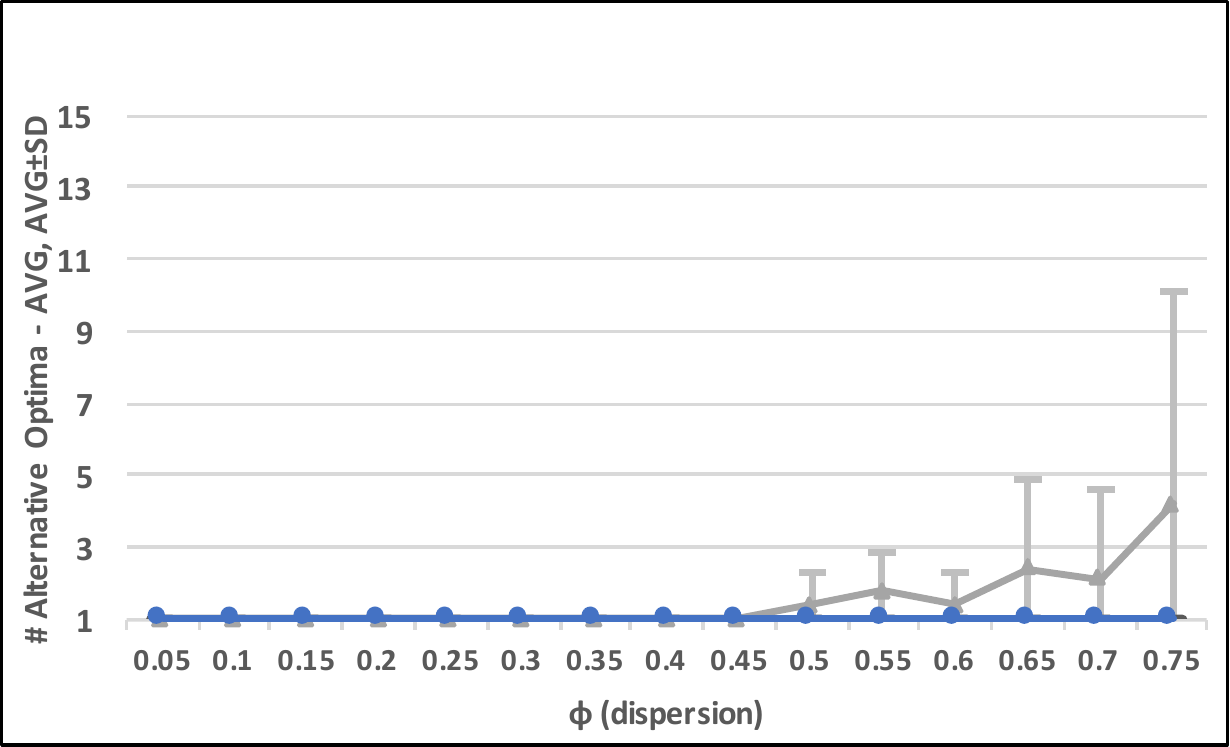}
		\caption{RIME2 Instances with $|V|=16$}
		\label{fig:rime2_16}
	\end{subfigure}
	\caption{$\tx$ evinces much less decisiveness than $\txS$ as the input rankings become noisier}
	\label{fig:simple}
\end{figure*}
The first set of instances is guided by the scenarios characterized in Table \ref{table:scenarios}. To obtain data with differing noise levels, each instance generates $K=100$ total rankings via RIME1 or RIME2 using single-dispersion values $\phi\in\{0.05,0.1,\dots,0.95,1.00\}$ (five $\phi$ values per group characterization in Table \ref{table:scenarios}); without loss of generality, the ground truth $\avt$ is set to $\mathbf{\epsilon}_n$ (the identity permutation) in all these instances. Prior to generating a set of input rankings $\{\av^k\}^K_{k=1}$, the object subset ranked by each $\av^k$, $V_{\av^k}\subseteq V$, is determined randomly along with its cardinality $|V_{\av^k}|\le n=|V|$, which is drawn from the uniform distribution $U(l, u)$. Different sizes of $V$ and $V_{\av^k}$ are tested for $k=1,\dots K$, namely $|V|=8$ with $|V_{\av^k}|\sim U(2,6)$, $|V|=12$ with $|V_{\av^k}|\sim U(3,9)$, and $|V|=16$ with $|V_{\av^k}|\sim U(4,12)$. These combinations are chosen to reflect realistic scenarios in which judges evaluate uneven but reasonable numbers of objects. For instance, it is unreasonable for most humans to objectively rank a large fraction of objects when $|V|=16$.

Figures \ref{fig:rime1_8}, \ref{fig:rime1_12}, \ref{fig:rime1_16} for RIME1 and Figures \ref{fig:rime2_8}, \ref{fig:rime2_12}, \ref{fig:rime2_16} for RIME2 summarize the results graphically (to distinguish the measures visually, $\tx$ lines are gray and thick and $\txS$ lines are blue and dotted). We point out that when $|V|\ge 16$ and $\phi>.75$, some instances could not finish solving within 24 hours or their B\&B trees exceeded memory. Therefore, the horizontal axes of Figures \ref{fig:rime1_16} and \ref{fig:rime2_16} stop at $\phi =.75$. These plots demonstrate that $\txS$ attained a lower AVG number of alternative optima than $\tx$ for nearly all of the tested $\phi$. When $\phi>.35$, $\tx$ AVG and SD values began to increase considerably. For example, they reached respective values of 42.2 and 115.7 in Figure \ref{fig:rime1_12} and 5.8 and 6.1 in Figure \ref{fig:rime2_12}. Conversely, $\txS$ AVG and SD values were never more than 2 and 2.5 in all the instances. Thus, while the ability of both measures to obtain a unique optimal solution decreased as $\phi$ and $|V|$ increased, the growth in magnitude and variability in the number of alternative optimal solutions was markedly less pronounced with $\txS$. \llV

\subsection{Assessing electoral fairness}\label{SS:Experiments_3}
As discussed in \S\ref{SS:tx_hat}, $\txS$ is designed to assign equitable voting power to each individual judge (i.e., to enforce electoral fairness) in the aggregation, whereas $\tx$ seems to provide increased representation to judges who evaluate more objects (see \S\ref{SS:taux_inadequate}). To gauge the impact of this fundamental difference, a new set of instances is constructed. In particular, Table \ref{table:scenarios} describes scenarios where more than one group participates, which is reflected by using multiple ground truths or ranges of $\phi$ to generate problem instances. 
\begin{table*}[ht]
\tiny \centering
	\caption{Complex scenarios constructed from multiple ground truths or dispersions}
	\label{table:comp_scenarios}
	\renewcommand{\arraystretch}{1.5}
	\begin{tabular}{|p{4.7cm}|p{7.25cm}|}
		\cline{1-2}\
		\bf Sampling parameters & \bf Generated instance description \\
		\cline{1-2}
		$(1\mi\alpha)\times100\%$ of input rankings generated with $(\avt, \phi_1 \in (0, 0.25])$ \mV\newline  $\alpha\times100\%$ of input rankings generated with $(\avt, \phi_2 \in (0.75, 1])$ & A [$(1\mi\alpha)\times100$]\%-majority hold nearly identical opinions, which are close to ground truth $\avt$, while a $[\alpha\times100$]\%-minority have nearly arbitrary opinions (i.e., with very low collective similarity), where $0<\alpha < .5$  \\ \cline{1-2}
		$(1\mi\alpha)\times100\%$ of input rankings generated with $(\avt, \phi \in (0, 0.25])$ \mV\newline $\alpha\times100\%$ of input rankings generated with $(\avt', \phi \in (0, 0.25])$ & A [$(1\mi\alpha)\times100$]\%-majority hold nearly identical opinions, which are close to ground truth $\avt$, while a $[\alpha\times100$]\%-minority seeks to distort the outcome through cohesive contrarian opinions (i.e., close to ground truth $\avt'$), where $0<\alpha < .5$.\\ \cline{1-2}
	\end{tabular}\llV
\end{table*}

The generated instances are guided by the Table \ref{table:comp_scenarios} characterizations. The first portion of these instances is generated with ground truth $\avt=\mathbf{\epsilon}_n$ and two dispersion parameters $\phi_1 \in \{0.05, 0.1, 0.15, 0.2, 0.25 \}$ and $\phi_2 \in \{0.8, 0.85, 0.9, 0.95, 1\}$ associated with a majority of experts and a minority of spammers (i.e., with jumbled heterogeneous opinions), respectively. The second portion of these instances is generated via a single dispersion parameter $\phi \in \{0.05, 0.1, 0.15, 0.2, 0.25\}$, which is used to draw both a majority of expert rankings with ground truth $\avt=\mathbf{\epsilon}_n$ and a minority of contrarian rankings with the reverse ground truth $\avt'$. Tested minority proportions are $\alpha\in\{0.05, 0.10, 0.15, 0.20\}$. For all of these instances we set $K=50$, $|V|=12$, and $|V_{\av^k}|\sim U(3,6)$ for $k=1,\dots,\lfloor(1\mi\alpha) K\rfloor$, where the latter is used to draw the number of objects ranked by the majority.  

The experiment tests two factors. First, it tests gradually increasing the uniform distribution minimum ($l$) and maximum $(u)$ parameters from 3 to 6 and from 6 to 9, respectively, for drawing the number of objects ranked by the minority. That is, the distribution of $|V_{\av^k}|$ changes from $U(3,6)$ to $U(6,9)$ for $k=\lfloor(1\mi\alpha) K\rfloor\ma1,\dots K$. Second, it tests the effect of composing the minority with spammers versus contrarians, where the minority proportions tested are $\alpha \in \{0.05, 0.10, 0.15, 0.20\}$. To test these factors, the experiment calculates the average Solution to Ground-truth Similarity (AVG SGS), i.e., the average $\txS$ correlation between each unique or alternative optimal solution and $\avt$. Note that  $\txS$ can be confidently used for this purpose since the aggregate rankings and the ground truth are complete---that is, $\tx$ and $\txS$ are interchangeable for this calculation (see \S\ref{SS:pairings}). Since the performance of $\tx$ was more unstable for RIME1 instances  (see Figure \ref{fig:simple}), the remaining experiments consider only RIME2 instances.

\begin{table}[ht!]
\begin{tabular}{p{0.03cm}cc}
                        & $\tau_x$ & $\hat{\tau}_x$ \\
\rotatebox{90}{\qquad \footnotesize{Spammers}} & \includegraphics[height=1.3in]{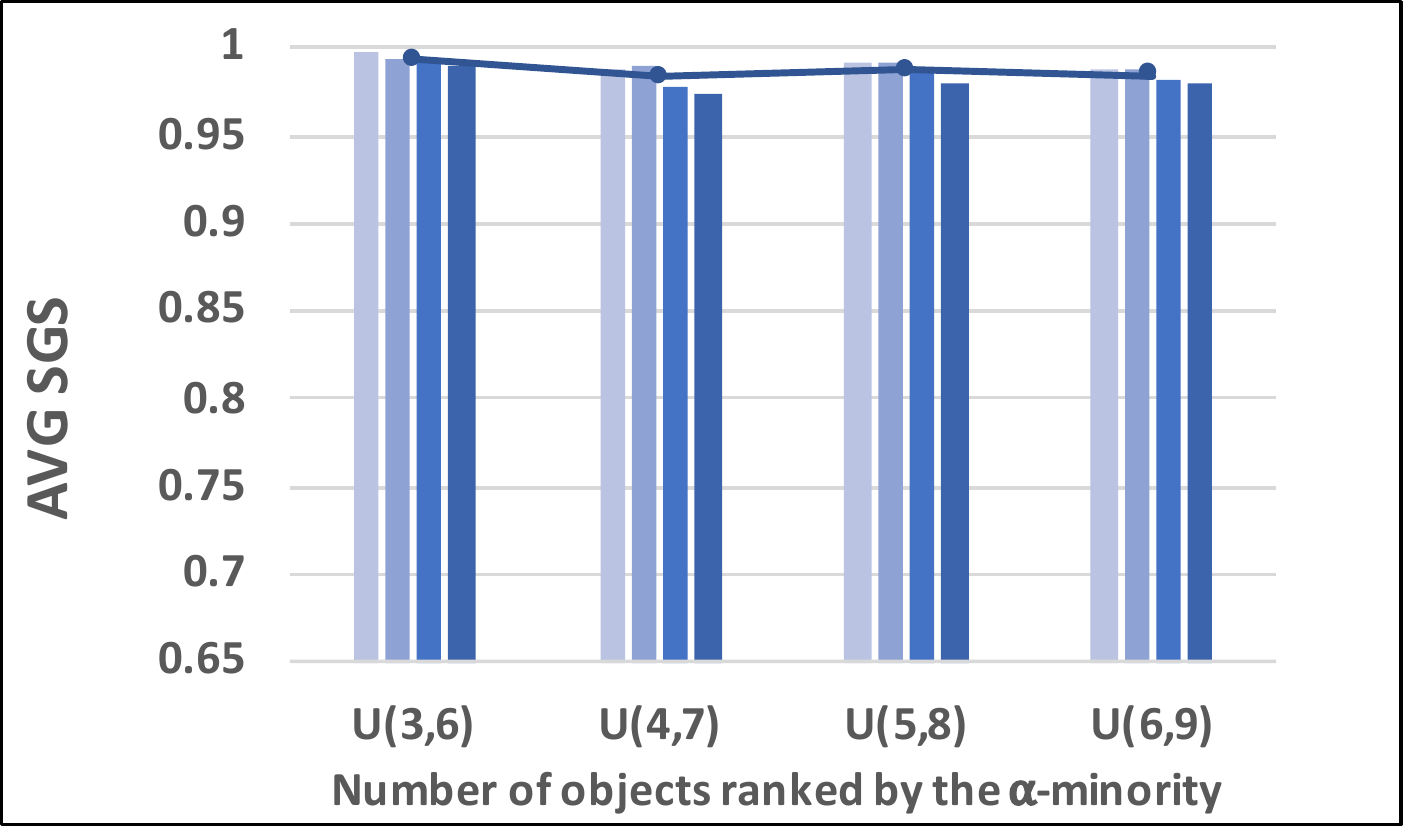} &\includegraphics[height=1.3in]{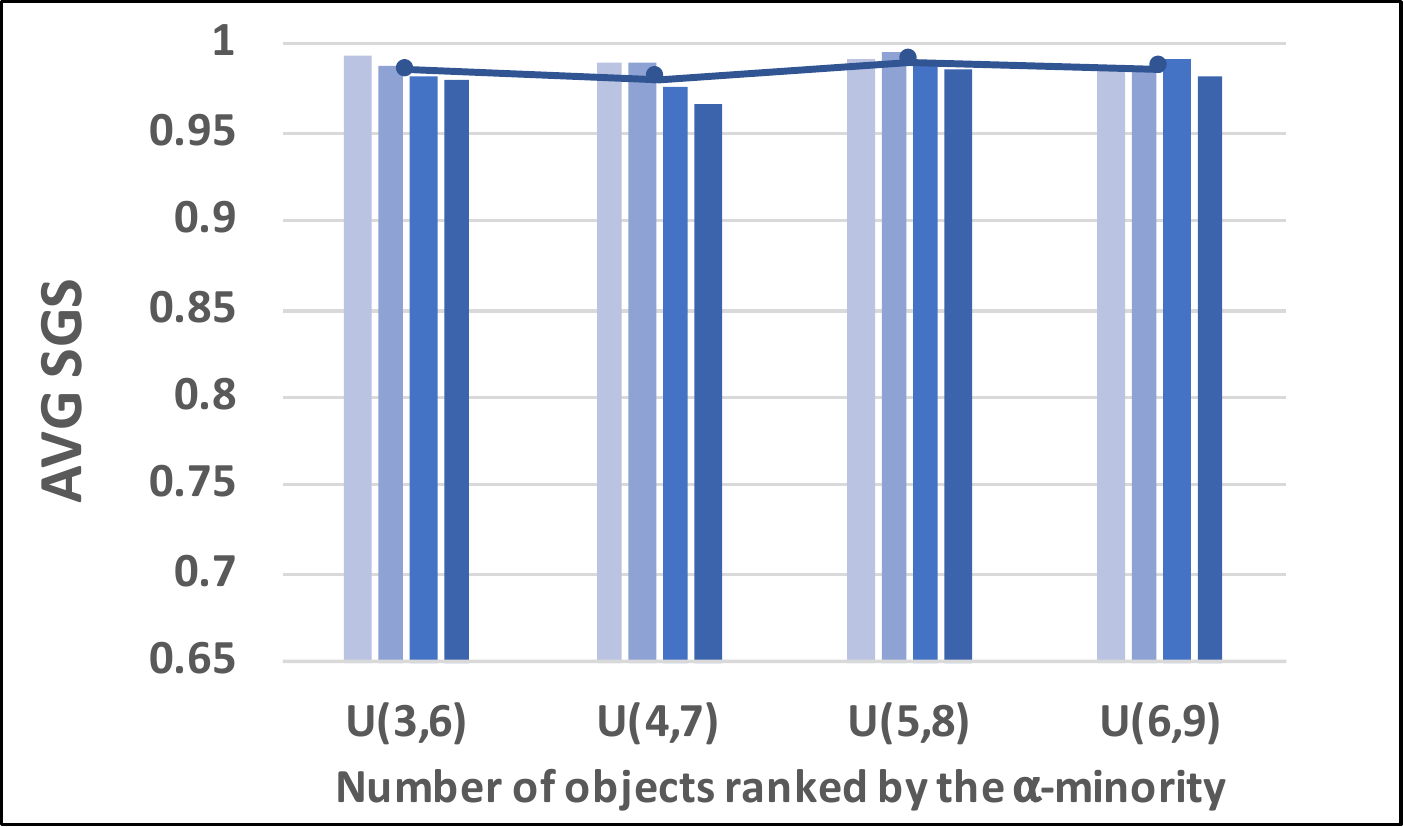} \\[-0.7ex]
&\multicolumn{2}{l}{\qquad \includegraphics[width=3.5in]{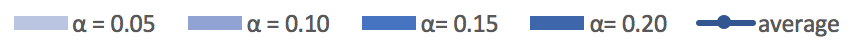}}\\[-0.7ex]
\rotatebox{90}{\qquad \footnotesize{Contrarians}} & \includegraphics[height=1.3in]{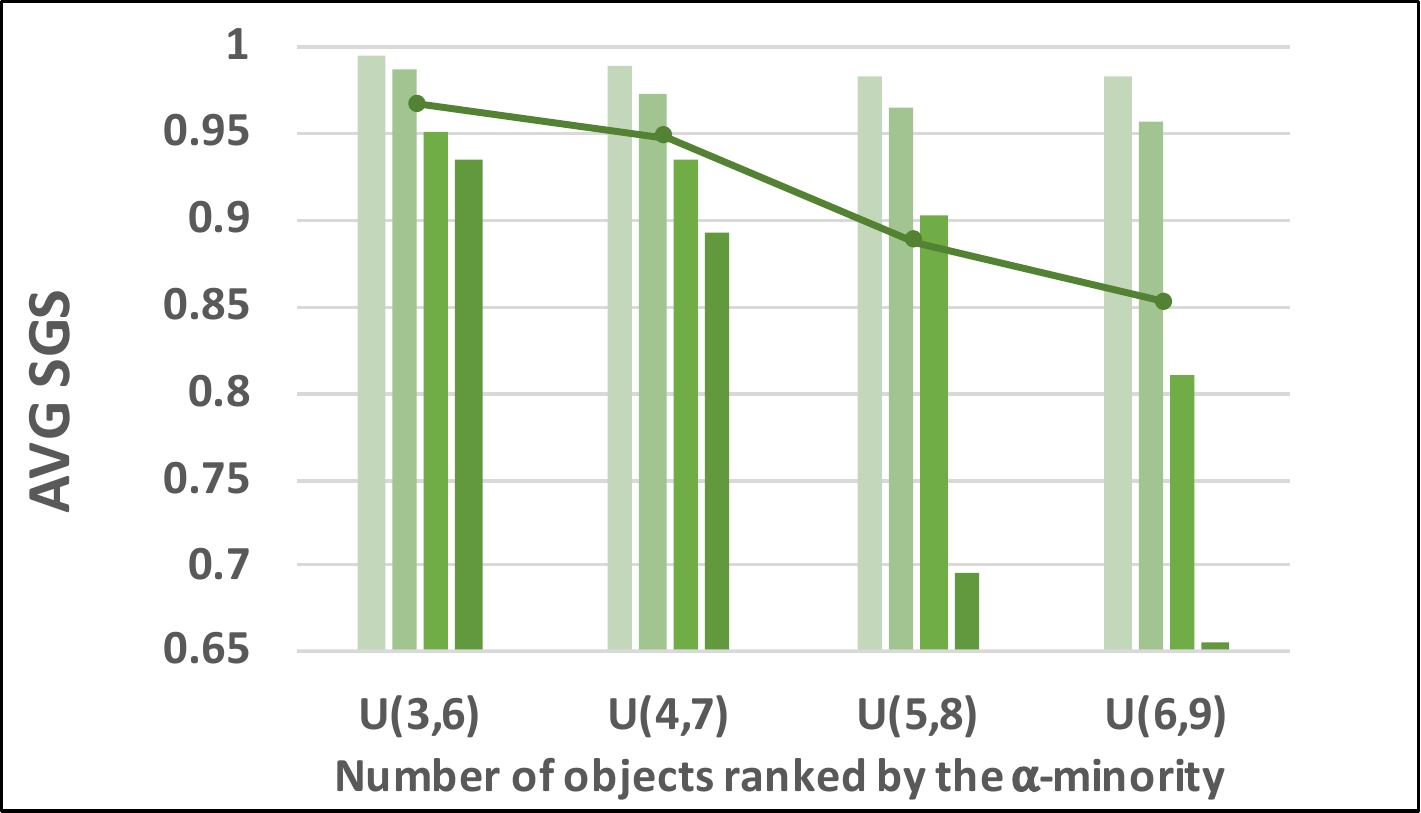} &\includegraphics[height=1.3in]{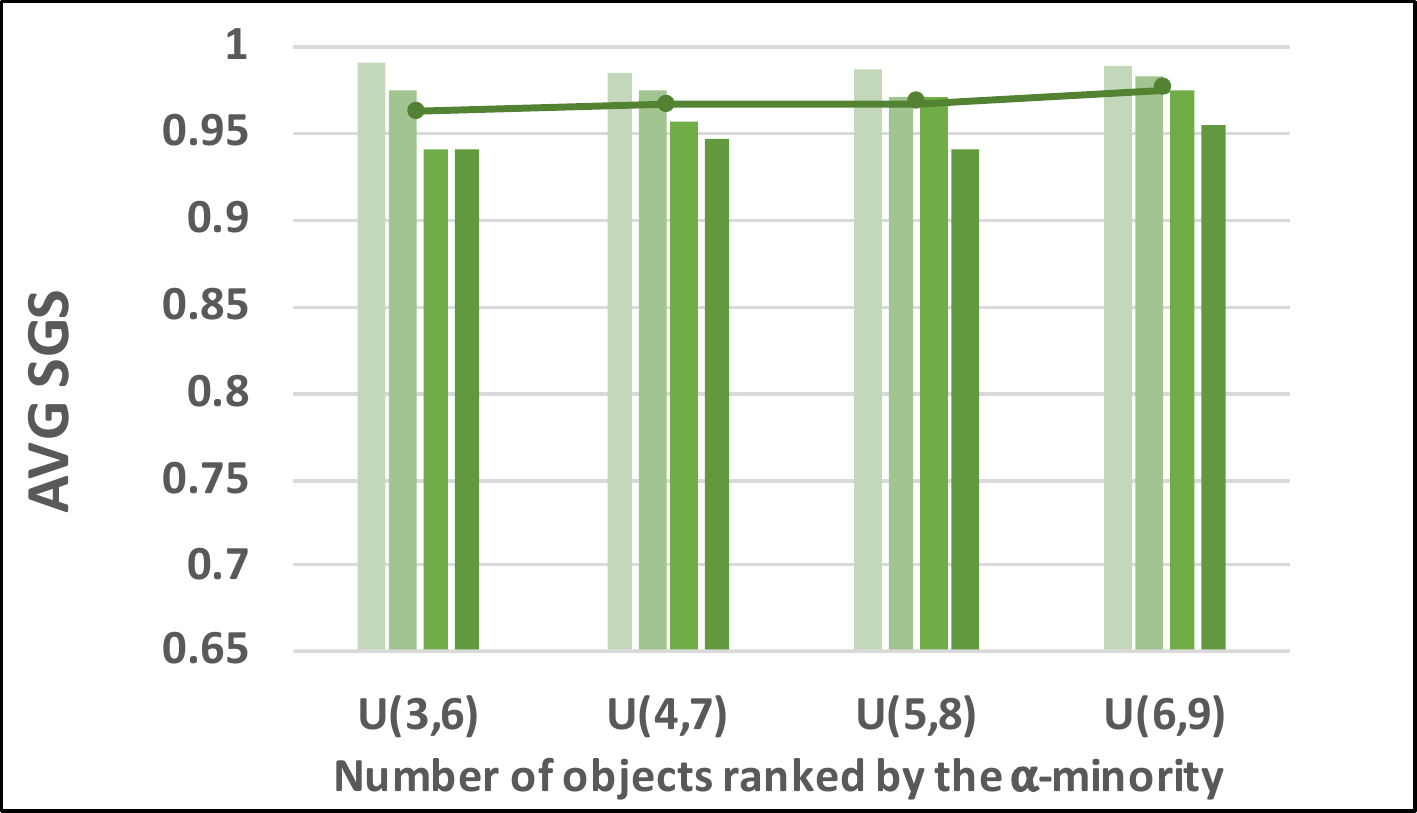}\\
&\multicolumn{2}{l}{\qquad\includegraphics[width=4in]{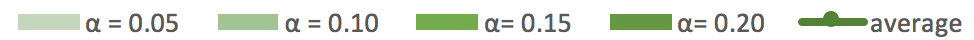}}
\end{tabular}\llV
\captionof{figure}{The similarity between the $\tx$ optimal rankings and the expert's ground truth  decreases sharply when a fixed minority of contrarians ranks more objects}\llV
\label{fig:spm_ctr_2}
\end{table}

Figure \ref{fig:spm_ctr_2} shows that the aggregate rankings obtained with $\tx$ yield smaller AVG SGS for minorities of contrarians. This occurs because  contrarians hold the reverse ground truth, which makes it difficult to get close to the majority's ground truth. Moreover, as the number of objects that contrarians evaluate increases, the average SGS decreases in Figure \ref{fig:spm_ctr_2}; in the worst case observed, AVG SGS becomes as low as 0.65, which occurs when $\alpha = 0.20$ and $U(6,9)$ for the number of objects ranked by the minority. Whereas a minority of contrarians results in much smaller average SGS values for $\tx$ compared with $\txS$, a minority of spammers causes a less pronounced difference. This is because spammers do not have strong cohesiveness among them, meaning they cannot affect the aggregate ranking as strongly. On the other hand, even if the percentage of contrarians is small, this causes the $\tx$ aggregate rankings to get further from the experts' ground truth. Thus, due to the implicit systematic bias induced by $\tx$, a higher number of objects ranked by a minority of contrarians can significantly affect fairness in the aggregation process as the resulting solutions get unreasonably far from the majority of experts' ground truth.

Notice that as the minimum and maximum parameters increase from $U(3,6)$ to $U(6,9)$ and the minority proportion remains fixed, AVG SGS decreases. In particular, the bottom-left of Figure \ref{fig:spm_ctr_2} shows that AVG SGS for $\tx$ decreased by more than 0.25 for a minority of contrarians with fixed $\alpha$. In contrast, the corresponding AVG SGS values for $\txS$ remained close to 1 and relatively stable in the bottom-right of Figure \ref{fig:spm_ctr_2}. This result indicates that $\txS$ is a more robust measure than $\tx$ for such types of nontrivial instances since it is not as adversely affected by judges who evaluate more alternatives.\llV 

\section{Discussion\llV}\label{Sec:Discussion}
This work makes several contributions to the area of robust ranking aggregation. Principally, it develops the $\txS$ ranking correlation coefficient, which fulfills the standard definitions of statistical correlation in the space of non-strict incomplete rankings. This ranking measure is applicable to situations where no assumptions should be made regarding individual preferences over unranked objects. Its formal derivation and axiomatic foundation ensure that $\txS$ assigns equitable voting power to each input ranking in the aggregation process, irrespective of the number of objectives it ranks. By also connecting $\txS$ with $\ksN$, this work enhances distance and correlation-based robust methodologies for ranking aggregation including the development of expedient optimization methodologies. The paper also develops a statistical sampling framework for generating non-trivial instances of the non-strict incomplete ranking aggregation problem.

In this research, ranking aggregation is applied to find the mathematical consensus between a set of subjective preferences. However, the ranking aggregation problem has wide-ranging applicability. In particular, the mathematical approach herein designed may be of use in numerous contexts where sets of incomplete ordinal data are compared to a ``gold standard'' \citep{yil08new} or aggregated to derive robust measures of central tendency \citep{dwo01ran}. For instance, the neutral treatment of incomplete rankings is relevant for the integration of omics-scale data in Biostatistics, where rank-based methods have been utilized due to their invariance to transformation and normalization \citep{lin10ran}. Within this context, it is often necessary to integrate ranked lists of genes arising from multiple technology platforms or studies, each of which may have access to overlapping yet differing parts of the genome. To avoid obtaining suboptimal results and introducing noise/bias to the aggregation, one must factor in the different underlying spaces of each list and ensure no assumptions are made about the genes outside each respective space \citep{lin10spa}.  For future work, we plan to extend and tailor the present framework to some of these applications. We will also analyze the properties of other plausible statistical distributions for sampling non-strict incomplete rankings. 

\newpage
\section{Appendix}\label{Sec:appendix}
\subsection{\bla Proofs of axiomatic foundation for $\txS$} \label{App:axioms_coefficient}
This subsection proves that the scaled Kendall-tau correlation coefficient $\txS$ satisfies Axiom 1-6 in \S\ref{SS:pairings}.\\

Axiom 1 (Relevance). The correlation discounts the unevaluated objects:\llV\lV
\begin{flalign*}
    \txS(\av,\bv)=\txS(\ap,\bp).
\end{flalign*}
\begin{proof}
From the definition of $\txS$ in Equation \eqref{eqn:ntaux}, the ranking-matrix elements for unevaluated objects are assigned to be 0 and, thus, the corresponding numerator terms $a_{ij}b_{ij}$ become 0. Therefore, it is valid to calculate the correlation coefficient by focusing on the mutually evaluated objects. \end{proof}

Axiom 2 (Commutativity). The correlation value is independent of the order in which  $\av$ and $\bv$ are compared:\llV\lV
\begin{flalign*}
    \txS(\av,\bv)=\txS(\bv,\av)
\end{flalign*}
\begin{proof}\renewcommand{\qedsymbol}{}
From the definition of $\txS$, we can write the following equations:
\begin{flalign*}
\txS(\av, \bv) &= \frac{\sum_{i=1}^{n} \sum_{j=1}^{n} a_{ij}b_{ij}}{\bar n(\bar n\mi1)}\\
&= \frac{\sum_{i=1}^{n} \sum_{j=1}^{n} b_{ij}a_{ij}}{\bar n(\bar n\mi1)}\\
&= \txS(\bv, \av) \qquad _\square
\end{flalign*}
\end{proof}

Axiom 3 (Neutrality) The correlation value is independent of the particular labeling of the objects:

If $\av'=\pi(\av)$ and $\bv'=\pi(\bv)$, then $\txS(\av,\bv)=\txS(\av',\bv')$, where $\pi$ is a permutation function such that $\pi: \{1,2,...,n\} \rightarrow \{1,2,...,n\}.$
\begin{proof}
Without loss of generality, assume that only two objects are permuted at a time, namely the $k$-th and $l$-th objects, with $k<l$. Let $A=[a_{ij}]$ and $B=[b_{ij}]$ be the pre-permutation ranking-matrices corresponding to ranking $\av$ and $\bv$, which are illustrated as:
\begin{equation*}\label{eq:appendrow}
  A = \left(\arraycolsep=1.4pt\def\arraystretch{0.9}\begin{array}{cc>{\columncolor{blue!20}}cc>{\columncolor{red!20}}ccc}
    a_{11} & \cdots & a_{1k} & \cdots & a_{1l} & \cdots & a_{1n} \\
    \vdots & \vdots  & \vdots & \cdots & \vdots & \cdots & \vdots \\
   \rowcolor{blue!20}
    a_{k1} & \cdots  & a_{kk} & \cdots & a_{kl} & \cdots & a_{kn} \\
    \vdots & \vdots  & \vdots & \cdots & \vdots & \cdots & \vdots \\
   \rowcolor{red!20}
    a_{l1} & \cdots  & a_{lk} & \cdots & a_{ll} &\cdots & a_{ln} \\
    \vdots & \vdots  & \vdots &  \cdots & \vdots &\cdots & \vdots \\
    a_{n1} &  \cdots  & a_{nk} & \cdots & a_{nl} &\cdots & a_{nn}\\
  \end{array}\right)
  \qquad B = \left(\arraycolsep=1.4pt\def\arraystretch{0.9}\begin{array}{cc>{\columncolor{blue!20}}cc>{\columncolor{red!20}}ccc}
    b_{11}  & \cdots & b_{1k} & \cdots & b_{1l} &  \cdots &b_{1n} \\
    \vdots   & \vdots  & \vdots & \cdots & \vdots & \cdots & \vdots \\
    \rowcolor{blue!20}
    b_{k1}   & \cdots  & b_{kk} & \cdots &  b_{kl} & \cdots & b_{kn} \\
    \vdots & \vdots  & \vdots & \cdots & \vdots & \cdots & \vdots \\
    \rowcolor{red!20}
    b_{l1}   & \cdots  & b_{lk} & \cdots &  b_{ll} & \cdots & b_{ln} \\
    \vdots & \vdots  & \vdots & \cdots & \vdots & \cdots & \vdots \\
    b_{n1}  &  \cdots  & b_{nk} & \cdots &  b_{nl} & \cdots & b_{nn} \\
  \end{array}\right).
\end{equation*}
 Let $\av'$ and $\bv'$ be the post-permutation rankings with corresponding ranking-matrices $A'=[a'_{ij}]$ and $B'=[b'_{ij}]$, which are illustrated as:
\begin{equation*}\label{eq:appendrow}
 A' = \left(\arraycolsep=1.4pt\def\arraystretch{0.9}\begin{array}{cc>{\columncolor{blue!20}}cc>{\columncolor{red!20}}ccc}
    a_{11} & \cdots & a_{1l} & \cdots & a_{1k} & \cdots & a_{1n} \\
    \vdots & \vdots  & \vdots & \cdots & \vdots & \cdots & \vdots \\
   \rowcolor{blue!20}
    a_{l1} & \cdots  & a_{ll} & \cdots & a_{lk} & \cdots & a_{ln} \\
    \vdots & \vdots  & \vdots & \cdots & \vdots & \cdots & \vdots \\
   \rowcolor{red!20}
    a_{k1} & \cdots  & a_{kl} & \cdots & a_{kk} &\cdots & a_{kn} \\
    \vdots & \vdots  & \vdots &  \cdots & \vdots &\cdots & \vdots \\
    a_{n1} &  \cdots  & a_{nl} & \cdots & a_{nk} &\cdots & a_{nn}\\
  \end{array}\right)
  \qquad B' = \left(\arraycolsep=1.4pt\def\arraystretch{0.9}\begin{array}{cc>{\columncolor{blue!20}}cc>{\columncolor{red!20}}ccc}
    b_{11}  & \cdots & b_{1l} & \cdots & b_{1k} &  \cdots &b_{1n} \\
    \vdots   & \vdots  & \vdots & \cdots & \vdots & \cdots & \vdots \\
    \rowcolor{blue!20}
    b_{l1}   & \cdots  & b_{ll} & \cdots &  b_{lk} & \cdots & b_{ln} \\
    \vdots & \vdots  & \vdots & \cdots & \vdots & \cdots & \vdots \\
    \rowcolor{red!20}
    b_{k1}   & \cdots  & b_{kl} & \cdots &  b_{kk} & \cdots & b_{kn} \\
    \vdots & \vdots  & \vdots & \cdots & \vdots & \cdots & \vdots \\
    b_{n1}  &  \cdots  & b_{nl} & \cdots &  b_{nk} & \cdots & b_{nn} \\
  \end{array}\right).
\end{equation*}
Note that unshaded elements in $A'$ and $B'$ remain the same after permutation (i.e., $a_{ij}=a'_{ij}$ for $i,j \neq k, l$). Since the $k$-th row (column) and $l$-th row (column) of $A$ and $B$ are exchanged, the remaining entries are given by:
 \begin{equation}\label{eqn:conversion}
 a'_{kk} = a_{ll}, \quad a'_{ll} = a_{kk}, \quad a'_{ik}=a_{il}, \quad a'_{il}=a_{ik}, \quad a'_{ki}=a_{li}, \quad a'_{li}=a_{ki}
\end{equation}
for every $i \neq k, l$ (the new entries for $\bv$ are defined similarly).
The permutation will affect only the numerator of $\txS(\av, \bv)$ (see Equation \eqref{eqn:taux_sm}). In particular, by using Expression \eqref{eqn:conversion}, the following equations can be derived:
{\small
\begin{align*}
   \sum_{i=1}^{n} \sum_{j=1}^{n} a_{ij}b_{ij} &=  \lH \sum_{i\neq k,l}^{n}  \sum_{j\neq k,l}^{n} \lH a_{ij}b_{ij} \lH + \lH \sum_{j\neq k,l} \lH (a_{kj}b_{kj}\lH  +\lH  a_{lj}b_{lj})\lH  + \lH \sum_{i \neq k,l}\lH (a_{ik}b_{ik}\lH  +\lH  a_{il}b_{il})\lH   + \lH \sum_{i=k,l}\sum_{j=k,l} a_{ij}b_{ij}\\
    &=\lH  \sum_{i\neq k,l}^{n} \sum_{j\neq k,l}^{n}\lH  a'_{ij}b'_{ij} \lH + \lH \sum_{j\neq k,l}\lH  (a'_{lj}b'_{lj} \lH +\lH  a'_{kj}b'_{kj})\lH  + \lH \sum_{i \neq k,l} \lH (a'_{il}b'_{il} \lH + \lH a'_{ik}b'_{ik})\lH  + \lH \sum_{i=k,l}\sum_{j=k,l} a'_{ji}b'_{ji} \\
    &=\lH  \sum_{i=1}^{n} \sum_{j=1}^{n} a'_{ij}b'_{ij}.
\end{align*}}
\noindent Therefore, $\txS(\av,\bv) = \txS(\av',\bv')$. Since any permutation can be described as a sequence of permutations of two objects at a time, $\txS(\av, \bv) = \txS(\av', \bv')$ holds for any permutation $\pi$.  
\end{proof}

Axiom 4 (Reduction). If $\av$ and $\bv$ agree except for a set $V'\subseteq V$, $\txS(\av,\bv)$ may be computed by focusing only on the objects in $V'$:  \llV
\begin{equation*}
    \txS(\av, \bv) = 1 + 2\txS(\av', \bv').
\end{equation*}
where $\av'=\av|_{V'}$ and $\bv'=\bv|_{V'}$, which are the rankings projected to the objects in $V'$.\bla

\begin{proof}\renewcommand{\qedsymbol}{}
By definition of the ranking-matrix representation of $\txS$, if rankings $\av$ and $\bv$ have positive agreement for objects $v_i, v_j$ (i.e., one prefers $v_i$ over $v_j$ and the other also prefers $v_i$ over $v_j$, or both tie $v_i$ and $v_j$), the corresponding numerator $a_{ij}b_{ij}$ becomes 1. Otherwise, $a_{ij}b_{ij}$ becomes -1. Let $p_{c}$ be the number of pairs in concordance and $p_{dc}$ be the number of pairs in discordance. Then, $\bar n(\bar n\mi1) = 2p_{c} + 2p_{dc}$ because there are two elements of ranking-matrix for $v_i$ and $v_j$, $a_{ij}$ and $a_{ji}$ ($b_{ij}$ and $b_{ji}$, respectively). The calculation of $\txS$ can be decomposed as follows:
\begin{equation*}\label{eqn:decomp_coeff}
\begin{aligned}
\frac{\sum_{i=1}^{n} \sum_{j=1}^{n} a_{ij}b_{ij}}{\bar n(\bar n\mi1)} = &\frac{1}{\bar n(\bar n\mi1)} \times 2p_{c}+\frac{-1}{\bar n(\bar n\mi1)} \times 2p_{dc}.
\end{aligned}
\end{equation*}
By replacing $2p_{c}$ with $\bar n(\bar n\mi1)-2p_{dc}$,
\begin{equation*}\label{eqn:decomp_coeff_agree}
\begin{aligned}
\frac{\sum_{i=1}^{n} \sum_{j=1}^{n} a_{ij}b_{ij}}{\bar n(\bar n\mi1)} &= \frac{1}{\bar n(\bar n\mi1)} \times (\bar n(\bar n\mi1)-2p_{dc})+\frac{-1}{\bar n(\bar n\mi1)} \times 2p_{dc}\\
&= 1 - \frac{2}{\bar n(\bar n\mi1)} \times 2p_{dc}\\
&= 1 + \frac{2\sum_{i=1}^{n} \sum_{j=1}^{n}a'_{ij}b'_{ij}}{\bar n(\bar n\mi1)}
\end{aligned}
\end{equation*} 
Hence, if $\av$ and $\bv$ agree except for a set $V' \subseteq V$, then $\txS$ can be calculated by focusing on the objects where $\av$ and $\bv$ disagree. That is,
\begin{equation*}
     \txS(\av, \bv) = 1 + 2\txS(\av', \bv') \qquad _\square
\end{equation*} 
\end{proof}

Axiom 5 (Relaxed Triangle Inequality).
Relationship among the three possible paired comparisons from three incomplete rankings: \begin{equation*}
    \txS(\av\pte,\bv\pte) + \txS(\bv\pte,\cv\pte) \leq \txS(\av\pte,\cv\pte) + 1; \llV
\end{equation*}
and equality holds if and only if $\bv\pte$ is between the other two projected rankings; here, $V_{\av,\bv,\cv} := \sA\inter \sB\inter\sC$ for concise representation. 
\begin{proof}
Let $\bar{n}=|V_{\av,\bv,\cv}|$. To investigate the relationship between $\txS(\av, \bv)$, $\txS(\bv, \cv)$, and $\txS(\av, \cv)$, begin by writing their corresponding definitions: 
\begin{flalign*}
\txS(\av,\bv) = \frac{\sum\limits_{i=1}^{n} \sum\limits_{j=1}^{n} a_{ij}b_{ij}}{\bar n(\bar n\mi1)},  \txS(\bv,\cv) = \frac{\sum\limits_{i=1}^{n} \sum\limits_{j=1}^{n} b_{ij}c_{ij}}{\bar n(\bar n\mi1)}, \txS(\av,\cv) = \frac{\sum\limits_{i=1}^{n} \sum\limits_{j=1}^{n} a_{ij}c_{ij}}{\bar n(\bar n\mi1)}.
\end{flalign*}
From these definitions, we can form the expression:
\begin{flalign*}
\txS(\av,\bv) + \txS(\bv,\cv) = \frac{\sum_{i=1}^{n} \sum_{j=1}^{n} b_{ij}(a_{ij} + c_{ij})}{\bar n(\bar n\mi1)}.
\end{flalign*}
There are three possibilities for the sum and product of $a_{ij}$ and $c_{ij}$:
\[ \begin{array}{lcll}
\mbox{1)  $a_{ij} = c_{ij} = 1$ } & \Longrightarrow & a_{ij} + c_{ij} = 2, & a_{ij}c_{ij} = 1 \\
\mbox{2) $a_{ij} = 1, c_{ij} = -1 \text{, or } a_{ij} = -1, c_{ij} = 1$} & \Longrightarrow &  a_{ij} + c_{ij} = 0, & a_{ij}c_{ij} = -1 \\
\mbox{3) $a_{ij} = c_{ij} = -1$} & \Longrightarrow &  a_{ij} + c_{ij} = -2, & a_{ij}c_{ij} = 1 \end{array}\] 

Now, when $\bv\pte$ is between the other two projected rankings (i.e., $b_{ij}$ is equal to either $a_{ij}$ or $c_{ij}$ or both, or $b_{ij}$ may also equal 1 when $a_{ij}$ and $b_{ij}$ disagree), $b_{ij}$ can be determined from the values of $a_{ij}$ and $c_{ij}$ as follows:
\[ \begin{array}{lcl}
\mbox{1)  $a_{ij} = c_{ij} = 1$ } & \Longrightarrow &  b_{ij} = 1 \\
\mbox{2) $a_{ij} = 1, c_{ij} = -1 \text{, or } a_{ij} = -1, c_{ij} = 1$} & \Longrightarrow &  b_{ij} = 1 \text{ or } -1\\
\mbox{3) $a_{ij} = c_{ij} = -1$} & \Longrightarrow &  b_{ij} = -1 \end{array}\]

Referencing the above cases, if $\bv\pte$ is between the other two projected rankings, the following equality always holds for each $i, j$:
\begin{equation}
b_{ij}(a_{ij}+c_{ij}) = a_{ij}c_{ij} + 1.
\end{equation}
Therefore, summing over all $i, j$ yields the following inequality:
\begin{flalign}\label{eqn:relaxed_equ}
\sum_{i=1}^{n} \sum_{j=1}^{n} b_{ij}(a_{ij}+c_{ij}) = \sum_{i=1}^{n} \sum_{j=1}^{n} a_{ij}c_{ij} + \bar{n}(\bar{n}-1).
\end{flalign}
By a similar analysis, if $\bv\pte$ is not between the other two projected rankings,
\begin{equation}
b_{ij}(a_{ij}+c_{ij}) < a_{ij}c_{ij} + 1,
\end{equation}
and summing over all $i, j$ yields the following inequality:   
\begin{flalign}\label{eqn:relaxed_ineq}
\sum_{i=1}^{n} \sum_{j=1}^{n} b_{ij}(a_{ij}+c_{ij}) < \sum_{i=1}^{n} \sum_{j=1}^{n} a_{ij}c_{ij} + \bar{n}(\bar{n}-1).
\end{flalign}
Combining equations \eqref{eqn:relaxed_equ} and \eqref{eqn:relaxed_ineq} yields the inequality:
\begin{flalign*}
\sum_{i=1}^{n} \sum_{j=1}^{n} b_{ij}(a_{ij}+c_{ij}) \leq \sum_{i=1}^{n} \sum_{j=1}^{n} a_{ij}c_{ij} + \bar{n}(\bar{n}-1).
\end{flalign*}
Therefore, dividing by $\bar n (\bar n\mi1)$, we obtain the desired expression.
\end{proof}

Axiom 6 (Scaling). The correlation range is between -1 and 1, inclusively:
\begin{equation*}\llV
-1\leq \txS(\av,\bv)\leq1;\llV 
\end{equation*}
with $\txS(\av,\bv)=1$ iff $\ap=\bp$ and $\txS(\av,\bv)=-1$ iff  $\bp$ is the reverse ranking of $\ap$ (the latter must be a linear ordering).
\begin{proof}
$(\Leftarrow)$ Assume $\ap$ and $\bp$ are the same ranking. Then,
\begin{flalign*}
\txS(\av,\bv) = \frac{\sum_{i=1}^{n}\sum_{j=1}^{n}a_{ij}b_{ij}}{\bar n(\bar n-1)} =  \frac{\sum_{i=1}^{n}\sum_{j=1}^{n}a_{ij}a_{ij}}{\bar n(\bar n-1)} = \frac{\bar n(\bar n-1)}{\bar n(\bar n-1)}=1.
\end{flalign*}
If $\bp$ is the reverse ranking of $\ap$ and $\bp$ and $\ap$ are linear orderings, then $b_{ij}$ always has the opposite value of $a_{ij}$. That is, $b_{ij}=-a_{ij}$, which leads to the following inequality: 
\begin{flalign*}
\txS(\av,\bv) = \frac{\sum_{i=1}^{n}\sum_{j=1}^{n}a_{ij}b_{ij}}{\bar n(\bar n-1)} =  \frac{\sum_{i=1}^{n}\sum_{j=1}^{n}a_{ij}(-a_{ij})}{\bar n(\bar n-1)} = \frac{-\bar n(\bar n-1)}{\bar n(\bar n-1)}=-1.
\end{flalign*}

$(\Rightarrow)$
Let $\txS(\av, \bv) = 1$. Then, $\sum_{i=1}^{n}\sum_{j=1}^{n}a_{ij}b_{ij}$ should be $\bar n(\bar n-1)$, which means that $a_{ij}b_{ij} = 1$ for every $v_i,v_j\in \sA\cap\sB$. That is, $\av$ and $\bv$ agree on all their preferences. On the other hand, to achieve $\txS(\av, \bv) = -1$, $\sum_{i=1}^{ n}\sum_{j=1}^{n}a_{ij}b_{ij}$ should equal $- \bar n(\bar n-1)$, which implies that $a_{ij}b_{ij} = -1$ for every $v_i,v_j\in \sA\cap\sB$ and that $\ap$ and $\bp$ are linear orderings. That is, $\ap$ and $\bp$ express opposing strict preferences over all object pairs. Therefore, if $\txS(\ap, \bp)=1$, $\ap$ and $\bp$ are the same ranking, and if $\txS(\ap, \bp)=-1$, $\bp$ is the reverse ranking of $\ap$ \bla.
\end{proof}

\subsection{Proof of theorems and corollaries} \label{App:proof_thm_cor}
\begin{customtheorem}{1}
(Linear transformation between $\txS$ and $\ksN$) Let $\av$ and $\bv$ be two arbitrary rankings over $n=|V|$ objects drawn from the space of non-strict incomplete rankings, $\Omega$. Then, the $\txS$ correlation coefficient and the $\ksN$ distance are connected through the following equation:
\begin{equation}
  \ksN(\av,\bv)=\frac{1}{2}-\frac{1}{2}\txS(\av,\bv).
\end{equation}
\end{customtheorem} 
\begin{proof}\renewcommand{\qedsymbol}{}
For succinctness, denote $\avh=\ap$ and $\bvh=\bp$ as the rankings over $\bar n\leq n$ objects obtained by projecting $\av$ and $\bv$ onto the subset of objects $\bar V=\AB$ ranked in common. Notice that $\avh$ and $\bvh$ are complete rankings over the same reduced universe of $\bar n$ objects (i.e., they lie in space $\Omega_C$ relative to $\bar V$). As such, using $1/2$ as the minimum $\ks$ distance unit, the corresponding $\tx$  and $\ks$ values for $\avh$ and $\bvh$ are equated as follows \citep{emo02new}:
\begin{equation*}
\tx(\avh,\bvh)=1-\frac{4\;\ks(\avh,\bvh)}{\bar n(\bar n\mi1)}, 
\end{equation*}
which expressed in terms of $\ks$ yields the equivalent relationship:
\begin{flalign}
\ks(\avh,\bvh)
              &=\frac{\bar n(\bar n\mi1)}{4}-\frac{\bar n(\bar n\mi1)\tx(\avh,\bvh)}{4}\\
              &=\frac{\bar n(\bar n\mi1)}{4}-\frac{\bar n(\bar n\mi1)\sum_{i=1}^{\bar n} \sum_{j=1}^{\bar n}\bar a_{ij}\bar b_{ij}}{4\bar n(\bar n\mi1)}\label{eqn:tauxn_to_npks2}\\
              &=\frac{\bar n(\bar n\mi1)}{4}-\frac{\sum_{i=1}^{n} \sum_{j=1}^{n} a_{ij} b_{ij}}{4},\label{eqn:tauxn_to_npks3}
\end{flalign}
where Equation \eqref{eqn:tauxn_to_npks2} applies the definition of $\tx$ (see Equation \eqref{eqn:taux}) with respect to $\avh$ and $\bvh$; and where Equation \eqref{eqn:tauxn_to_npks3} cancels a common factor in the second term and utilizes the fact that unranked items in either ranking vector contribute nothing to the sum---that is the matrix inner products are identical in the original and projected spaces. Now, multiplying both sides of Equation \eqref{eqn:tauxn_to_npks3} by $[\bar n(\bar n\mi1)/2]^{-1}$ gives:
\begin{flalign*}
\frac{\ks(\avh,\bvh)}{\bar n(\bar n\mi1)/2}&=\frac{1}{2} -\frac{\sum_{i=1}^{n} \sum_{j=1}^{n} a_{ij} b_{ij}}{2\bar n(\bar n\mi1)}\\
\Rightarrow \mmmH\ksN(\av,\bv)&=\frac{1}{2} -\frac{1}{2}\txS(\av,\bv) \qquad _\square
\end{flalign*}
\end{proof}

\begin{customtheorem}{2}
(Linear transformation between $\tx$ and $\ksP$) \textit{Let $\av$ and $\bv$ be two arbitrary rankings of $n=|V|$ objects drawn from the space of non-strict incomplete rankings, $\Omega$. Then, the $\tx$ correlation coefficient and the $\ksP$ distance are connected through the following equation:
\begin{equation*}
  \ksP(\av,\bv)=\frac{\bar n(\bar n\mi1)}{4}-\frac{n(n-1)}{4}\tx(\av,\bv)
\end{equation*}
where $\bar n=|\bar V|=|\AB|$ (i.e., the number of objects explicitly ranked by both $\av$ and $\bv$).}
\end{customtheorem}
\begin{proof}
  From Theorem \ref{thm:txS_to_ksN}, we have that:
  \[\ksN(\av,\bv)=\frac{1}{2}-\frac{1}{2}\txS(\av,\bv),\]
  which can be expanded via Equations \eqref{eqn:npks} and \eqref{eqn:ntaux} as:
  \[\frac{\ks(\ap,\bp)}{\bar n(\bar n-1)/2}=\frac{1}{2}-\frac{\sum_{i=1}^{n}\sum_{j=1}^{n}a_{ij}b_{ij}}{2\bar n(\bar n-1)}.\]
  Thus, multiplying both sides by $\bar n(\bar n\mi1)/2$ yields:
  \begin{flalign*}
  \ksP(\av,\bv) &=\frac{\bar n(\bar n\mi1)}{4}-\frac{1}{4}\sum_{i=1}^{n}\sum_{j=1}^{n}a_{ij}b_{ij}\\
  &=\frac{\bar n(\bar n\mi1)}{4}-\frac{n(n-1)}{4}\left[\frac{\sum_{i=1}^{n}\sum_{j=1}^{n}a_{ij}b_{ij}}{n(n-1)}\right]
  \end{flalign*}
  which completes the proof since the bracketed expression matches the definition of $\tx(\av,\bv)$. 
\end{proof}

\begin{customcorollary}{1}
The respective NIRA optimization problems typified by $\txS$ and $\ksN$ are equivalent and, thus, provide identical consensus rankings. Similarly, the respective NIRA optimization problems typified by $\tx$ and $\ksP$ are equivalent and, thus, provide identical consensus rankings.
\end{customcorollary}

\proof The first part of corollary is established through the following series of equations:\llV
\begin{flalign}
\underset{r\in\Omega_C}{\arg\min} \sum_{k=1}^{K}\ksN(\rv,\av^k)&=\underset{r\in\Omega_C}{\arg\max} \sum_{k=1}^{K}-\ksN(\rv,\av^k)\label{eqn:minDis_to_maxAgr1}\\
              &=\underset{r\in\Omega_C}{\arg\max} \sum_{k=1}^{K}-\left[\frac{1}{2}-\frac{1}{2}\txS(\rv,\av^k)\right]\label{eqn:minDis_to_maxAgr2}\\
              &=\underset{r\in\Omega_C}{\arg\max} \sum_{k=1}^{K}\txS(\rv,\av^k),\label{eqn:minDis_to_maxAgr3}
\end{flalign}
where the last equation results from the fact that scalars common to every term in the sum and constant terms do not impact the optimal solution. 

Similarly, the second part of the corollary can be proved via the following series of equations:
\begin{flalign}
&\mmH\mH\phantom{=}\underset{r\in\Omega_C}{\arg\min} \sum_{k=1}^{K}\ksP(\rv,\av^k)\\
               &=\underset{r\in\Omega_C}{\arg\max}\sum_{k=1}^{K}-\ksP(\rv,\av^k)\label{eqn:minDis_to_maxAgr1}\\
              &=\underset{r\in\Omega_C}{\arg\max} \sum_{k=1}^{K}- \Bigl[\frac{(|V_{\rv}\cap V_{\av^k}|)(|V_{\rv}\cap V_{\av^k}|\mi1)}{4}-\frac{n(n\mi1)}{4}\tx(\rv,\av^k)\Bigr]\label{eqn:minDis_to_maxAgr2}\\
              &=\underset{r\in\Omega_C}{\arg\max} \sum_{k=1}^{K}\frac{n(n\mi1)}{4}\tx(\rv,\av^k)-\frac{(| V_{\av^k}|)(|V_{\av^k}|\text{-}1)}{4}\label{eqn:minDis_to_maxAgr3}\\
             & =\underset{r\in\Omega_C}{\arg\max} \sum_{k=1}^{K}\tx(\rv,\av^k)\label{eqn:minDis_to_maxAgr4}
\end{flalign}
where Equation \eqref{eqn:minDis_to_maxAgr2} ensues from Theorem \ref{thm:tx_to_ksP}; where Equation \eqref{eqn:minDis_to_maxAgr3} results from the fact that, since $r$ must be a complete ranking, $|V_{\rv}\cap V_{\av^k}|=|V_{\av^k}|$ for every $k$; and, where Equation \eqref{eqn:minDis_to_maxAgr4} results from the fact that scalars common to every term in the sum as well as constant terms (i.e, the second term in Equation \eqref{eqn:minDis_to_maxAgr3} is independent of any candidate solution) have no bearing on the optimal solution.
\eproof

\begin{customcorollary}{2}
\bla The correlation-based NIRA is $\cal{NP}$-hard.
\end{customcorollary}
\begin{proof}
The distance-based NIRA was proven to be $\cal{NP}$-hard in  \cite{mor16axi}. Since solving the correlation-based NIRA is equivalent to solving the distance-based NIRA by Corollary \ref{cor:unbiasedEquiv}, the former problem is also  $\cal{NP}$-hard.
\end{proof}

\begin{customtheorem}{3}
(Succinct function of cumulative agreement for $\txS$) Let $\rv\in\Omega_C$, $\av^k\in\Omega$, and $\bar n^k = |V_{\av^k}|$ (the number of objects ranked by $\av^k$), for $k=1,\dots,K$. Then, the $\txS$ cumulative correlation between $\rv$ and $\{\av^k\}_{k=1}^K$ can be computed according to the function:
\begin{equation*}
      \sum_{k=1}^{K}\txS(\rv, \av^k)=
\sum_{i=1}^{n}\sum_{j=1}^{n}\hat A_{ij}r_{ij},
\end{equation*}
where $[\hat A_{ij}]=\sum_{k=1}^{K}\frac{a^k_{ij}}{\bar n^k(\bar n^k-1)}$ is the \textit{scaled combined ranking-matrix (SCR)}.
\end{customtheorem}
\begin{proof}\renewcommand{\qedsymbol}{}
Since $\rv\in\Omega_C$, the term $\txS(\rv,\av^k)$ can be simplified as follows:
\begin{equation*}
   \txS(\rv,\av^k) = \frac{\sum_{i=1}^{n}\sum_{j=1}^{n}a^k_{ij}r_{ij}}{\abs{V_{\av^k}\cap V}(\abs{V_{\av^k}\cap V}-1)} = \sum_{i=1}^{n}\sum_{j=1}^{n}\frac{a^k_{ij}r_{ij}}{\bar n^k(\bar n^k-1)}.
\end{equation*}
Thus, the denominator associated with each term is constant irrespective of the candidate-solution vector, thereby yielding the equivalent expressions:
\begin{equation*}
      \sum_{k=1}^{K}\tx(\rv,\av^k)=
\sum_{k=1}^{K}\sum_{i=1}^{n}\sum_{j=1}^{n}\frac{a^k_{ij}}{\bar n^k(\bar n^k-1)}r_{ij}=\sum_{i=1}^{n}\sum_{j=1}^{n}\hat A_{ij}{r_{ij}}.  \qquad _\square
\end{equation*}
\end{proof}

\subsection{A flowchart and description of the branch and bound algorithm}\label{App:b&b}
Before introducing the branch and bound algorithm (B\&B), it is necessary to discuss the concept of an  object-ordering and its connection to a ranking. An  \textit{object-ordering}  $\av^{\mi1}$ is induced by a mapping $\Psi:\av\in\{1,\dots,n\}^n \rightarrow \av^{-1}\in W(\{1,\dots,n\})$, where $W(\{1,\dots,n\})$ denotes the set of weak orders (or complete preorders) on $n$ objects. That is, $\Psi(\av)$ sorts the objects in $V_{\av}$ from best to worst, according to their ranks in $\av$. For example, for $\av=(1,5,2,4,3)$, $\Psi(\av) = \av^{\mi1} = (v_1,v_3,v_5,v_4,v_2)$. Extending this notation, $\av^{\mi1}(i)$ specifies the $i$th-highest ranked object in $\av$, when $\av$ does not contain ties \citep{doi04rep}---that is, $\Psi$ is a bijection in this case and the inverse function $\Psi^{-1}$ returns a linear order. When $\av$ contains ties, $\Psi$ sorts the objects into preference equivalence classes.  For example, for $\av=(1,3,3,1,5)$, $\Psi(\av)=\av^{\mi1}=(\langle v_1,v_4 \rangle ,\langle v_2,v_3 \rangle,v_5)$. In the case of ties, the inverse mapping $\Psi^{-1}(\av^{\mi1})$ returns the ranking obtained by labeling each object with its corresponding equivalence class position in $\av^{\mi1}$. For example, for $\av^{\mi1}=(v_1,\langle v_2,v_4 \rangle, v_5,v_3)$, $\Psi^{-1}(\av^{\mi1})=\av=(1,2,5,2,4)$.

\begin{figure*}[h!]\noindent\centering\makebox[\textwidth]
	{\includegraphics[width=6.5in]{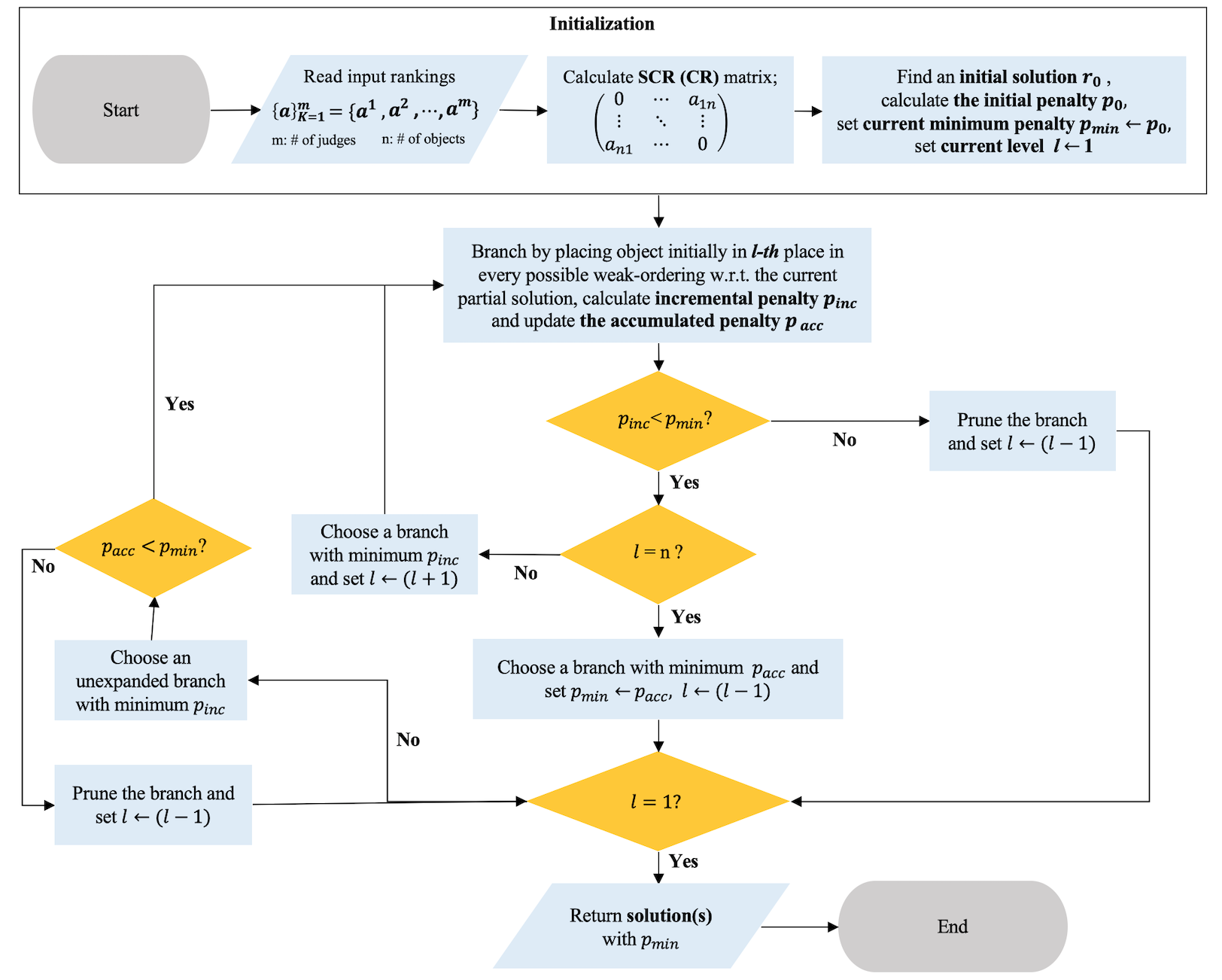}}
\end{figure*}
B\&B is applied as follows. First, the absolute values of the SCR (CR, respectively) matrix entries are summed to yield an upper bound on the cumulative correlation achievable by any candidate-solution vector \cite{emo02new}. An initial deviation penalty corresponding to a user-specified starting solution $\rv_0\in\Omega_C$ (obtained randomly or via a heuristic) is then calculated by subtracting its objective value from said upper bound. To describe the ensuing steps, recall that $\rv_0^{-1}$ is the object-ordering induced by the mapping function $\Psi(\rv_0)$. For $i=2,\dots,n$, the algorithm calculates incremental penalties of fixing object $\rvI{i}$ (ranked $i$th in the reference starting solution) to every possible pairwise preference relative to a candidate sub-ranking of objects $\rvI{1},\dots,\rvI{i\mi1}$ by inspecting the respective SCR entries. Three branches are created to reflect the possible ordinal relationships---i.e., preferred, tied, and dispreferred---between $\rvI{i}$ and each $\rvI{j}\in\{\rvI{1},\dots,\rvI{i\mi1}\}$. If the incremental penalty of a branch exceeds the current minimum penalty, the branch is pruned; otherwise it is explored by considering the next object, $\rvI{i\ma1}$. B\&B prioritizes newly created branches when there are multiple branches to explore. Once a complete ranking is obtained, the minimum penalty is updated and the ranking is saved as a possible solution; at the end of the algorithm, all rankings with the final minimum penalty are returned as the set of optimal solutions (i.e., the median or consensus rankings).

\subsection{\bla Exact integer programming formulation}\label{App:IP}
\citeauthor{yoo2018newinteger} \citep{yoo2018newinteger} recently derived the following set of linear constraints that guarantee that a matrix $R\in\{-1,0,1\}^{n\times n}$ is a valid ranking-matrix according to Equation \eqref{eqn:taux_sm}: 
\begin{subequations}
\begin{align}
r_{ij} - r_{kj} - r_{ik} \geq -1 \quad & i, j, k = 1,...,n; \quad i \neq j \neq k \neq i \label{eqn:const}\\ 
r_{ij} + r_{ji} \geq 0 \quad & i, j = 1,...,n;  \quad i \neq j \label{eqn:negative}\\  
r_{ii} = 0 \quad & i = 1,...,n; \label{eqn:diagonal}\\
r_{ij} - 2y_{ij} = -1 \quad & i, j = 1,...,n;  \quad i \neq j \label{eqn:aux} \\
r_{ij} \in \mathbb{Z}, y_{ij} \in \mathbb{B} \quad & i, j = 1,...,n.
\label{eqn:integer}
\end{align}
\end{subequations}
Constraint \eqref{eqn:const} eliminates the occurrence of pairwise preference cycles. Constraint \eqref{eqn:negative} restricts at least one of $r_{ij}$ and $r_{ji}$ to be positive since $r_{ij}$ and $r_{ji}$ cannot both be negative. The diagonal elements must be set to 0, which is represented by constraint \eqref{eqn:diagonal}. The off-diagonal elements must be non-zero values, specifically, they must be equal to 1 or -1. This is enforced via auxiliary binary variable $y_{ij}$ in constraint \eqref{eqn:aux}. Constraint \eqref{eqn:integer} states the domains of $r_{ij}$ and $y_{ij}$. Using this constraint set, the correlation-based NIRA for $\txS$ is obtained by appending the following objective function:\llV
\begin{equation*}
\arg\max_{\rv\in\Omega_C}\sum_{k=1}^{K}\txS(\rv,\av^k)
  =\arg\max_{\rv\in\Omega_C}\sum_{i=1}^{n}\sum_{j=1}^{n}\hat A_{ij}r_{ij}
\end{equation*}\bla 
which uses the SCR matrix $[\hat A_{ij}]$ defined in Equation \eqref{eqn:txSCI}. Note that the respective solution ranking $\rv\in\Omega_C$ is obtained by sorting the row sums (or column sums) of $[r_{ij}]$ in non-increasing (non-decreasing, resp.) order.

\newpage

\bibliography{mybibfile.bib}

\pagebreak

\end{document}